\documentclass{issue-lpr}
\usepackage[english]{babel}
\usepackage{times}
\usepackage{color}
\newcommand{\ket}[1]{| #1 \rangle}
\newcommand{\bra}[1]{\langle #1 |}
\newcommand{\braket}[2]{\langle #1 | #2 \rangle}

\newcommand{\nvminus}{NV$^-$~}
\newcommand{\tripletA}{$^3A$~}
\newcommand{\tripletE}{$^3E$~}
\newcommand{\singletA}{$^1A$~}

\newcommand{\cthirteen}{$^{13}$C~}
\newcommand{\ctwelve}{$^{12}$C}

\graphicspath{{./}{figs/}}

\setpages{5}
\setvolume[1]{1}

\setyear{2008}%

\setdoi{20070001}%

\begin{document}

\titlefigure[width=0.9 \columnwidth]{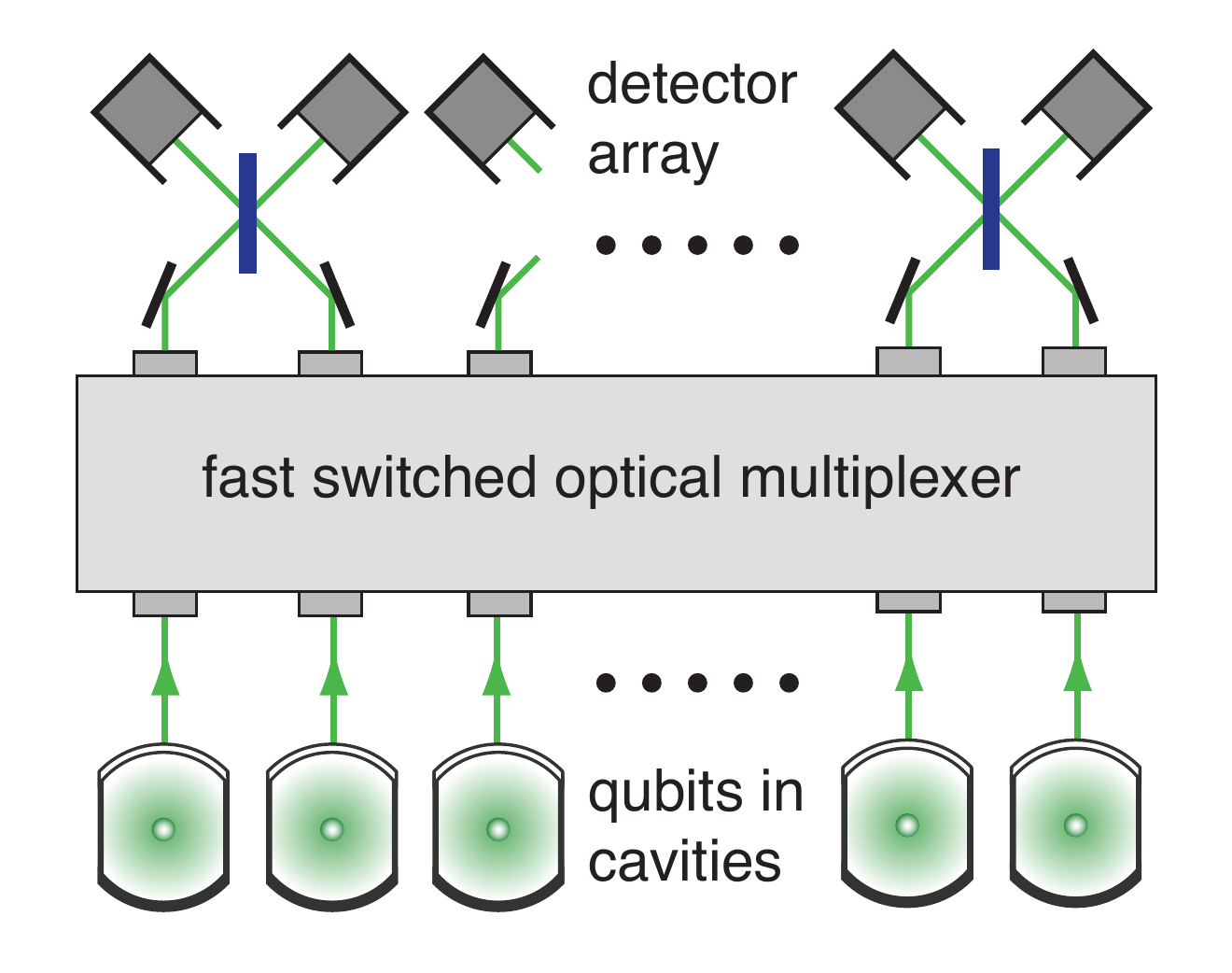}
\titlefigurecaption{Schematic architecture for a measurement-based quantum computer.}

%%% Give an abstract text:
\abstract{This article aims to review the developments, both theoretical and experimental, that have in the past decade laid the ground for a new approach to solid state quantum computing. Measurement-based quantum computing (MBQC) requires neither direct interaction between qubits nor even what would be considered controlled generation of entanglement. Rather it can be achieved using entanglement that is generated probabilistically by the collapse of quantum states upon measurement. Single electronic spins in solids make suitable qubits for such an approach, offering long coherence times and well defined routes to optical measurement. We will review the theoretical basis of MBQC and experimental data for two frontrunner candidate qubits -- nitrogen-vacancy (NV) centres in diamond and semiconductor quantum dots --  and discuss the prospects and challenges that lie ahead in realising MBQC in the solid state.}

%%% In the following lines, title and authoring informaton are to be 
%%% given.  First the article's main title: 
\title{Prospects for measurement-based quantum computing with solid state spins}

%%% Please name all authors and tag them with their respective 
%%% institute(s) using the \inst directive.  The corresponding author 
%%% gets an additional asterisk.
\author{Simon C. Benjamin,$^{1,2}$ Brendon W. Lovett,$^1$ and Jason M. Smith~$^1$}

%%% Here the institutes are to be given.  If there is more than one, 
%%% the entries have to be separated by \and tags.
\institute{%
University of Oxford, Department of Materials, Parks Road, Oxford OX1 3PH, UK \and Centre for Quantum Technologies, National University of Singapore, 3 Science Drive 2, 117543 Singapore
}

%%% Give the email address of the corresponding author:
\mail{e-mail: jason.smith@materials.ox.ac.uk}

%%% If the article has a very long title or lots of authors, then 
%%% please give shorter versions of that for the column title:
\titlerunning{MBQC using solid state spins}
\authorrunning{Benjamin, Lovett, and Smith}
%%% If there are more than two authors, please indicate that in the 
%%% form ``F. Author et al.''

%%% Give some keywords and PACS code(s) for the article, if you have:

%%% The following will be completed by the publisher: 
\received{\ldots}
\published{\ldots}
\maketitle
\markboth{thedoi}{thedoi}
\tableofcontents

%%% Here, the main text starts.

\section{Introduction}
When quantum computing was first conceived, it was generally assumed that to achieve a universal set of quantum gates one would need to be able to generate entanglement between qubits in a deterministic and reversible fashion. Early experimental architectures with electronic qubits therefore focused on systems in which the qubits were spaced closely enough, within tens of nanometres or less, to communicate with each other through some local interaction that could be gated using external controls \cite{Kane98,Loss98}. Despite some impressive achievements and beautiful science resulting from this work, the need to locate qubits so closely to each other whilst retaining full control on them individually presents enormous challenges, especially in the design of scalable architectures.

In the past decade, theoretical schemes for `measurement - based', or `one-way' quantum computing (MBQC) have provided a promising alternative to the circuit-based model. Rather than entanglement production being part of a quantum algorithm, an entangled state is generated in advance, and the algorithm is then executed via measurements which consume that entanglement. Such an approach brings the major advantage that errors in entanglement generation can be identified and corrected without prejudicing the algorithm. This leads directly to the realisation that entanglement generation can be probabilistic (provided that successes and failures are heralded) and therefore that entanglement can be generated by `measuring out' terms in the product state of two or more uncorrelated systems. The qubits can be spatially remote, provided we are able to construct an appropriate measurement. 

The first demonstration of measurement-based entanglement of macroscopically separate systems was reported by Chou et al using atomic ensembles in 2005 \cite{Chou05}, and entanglement of two remote single atoms was reported by Moehring et al in 2007 \cite{Moehring07}. Such experiments will almost certainly be extended to entanglement of multiple qubits in these very promising systems and could pave the way towards fully scalable quantum computing. Moreover, in this review we shall argue that the ability to use remote qubits is a particular boost to efforts to realize quantum computing in {\em solid state} systems. Not only does it solve the problem of single qubit manipulation outlined above, it also allows us to address such issues as inhomogeneities in the qubit local environments and low yields of working qubits. To build a ten qubit computer it is no longer necessary to be able to fabricate ten working qubits on a single device - one can fabricate ten `devices' each with one working qubit. The idea of a distributed quantum computer was thus born (see title figure), in which discrete modules containing individual qubits act as nodes connected to an optical multiplexing system. The multiplexer and qubit modules are all controlled by a classical computer that makes real time decisions based on the pattern of detection events and executes subsequent operations accordingly - a process known as `feed forward decision making'. The architecture is inherently scalable by the addition of further nodes and a larger multiplexer, provided (as with any quantum computation scheme) all the necessary operations can be executed within the coherence time of the qubits.

Even in this alternative measurement based paradigm for performing quantum computing, DiVincenzo's criteria~\cite{divincenzo00} of long coherence times, controlled qubit initialisation and manipulation, and efficient optical readout are principal considerations. Electron spins in solids can offer all of these, and in this review we will discuss two leading contenders, namely nitrogen vacancy defects in diamond and Stranski Krastanow quantum dots.

The paper is laid out as follows. In Section \ref{theory} we first review the basic physics and mathematics behind measurement - based quantum computing, highlighting the aspects that lead to practical requirements that differ from those of circuit-based schemes. In Section \ref{spins} we then discuss the physics that is particular to electron spin qubits in solids, such as decoherence due to interactions with the lattice and other neighbouring spins, and the requirements placed on the qubit modules by the need to perform high fidelity optical measurement of the spins. Section \ref{qubits} then focuses on our two candidate qubits, the properties that make them attractive for MBQC, and highlights the strengths and weaknesses of each. We conclude by providing an outlook for the challenges that remain in realising measurement-based quantum computing in the solid state.

\section{Theory of measurement-based quantum computing}
\label{theory}

In 1999 and 2001, two important theoretical insights into the creation and exploitation of entanglement were attained. The first was the discovery of an elegant method of entanglement generation by exploiting the act of measurement~\cite{Cabrillo99}. The second insight was that certain many-qubit entangled states constitute an {\em enabling resource} for quantum computing, in the sense that if one were given such a resource of sufficient size then one could perform any chosen algorithm just by measuring the qubits one at a time~\cite{RausBriegel01}. Taken together, these two ideas allow us to see that measurement can be the {\em sole} driving force of a quantum computation, and we are lead to consider computer architectures that are very different to the early circuit-based concepts such as Kane's proposal~\cite{Kane98}.

Here we will briefly review these two advances, and we will consider some schemes for computation that are tenable in light of these insights.

\subsection{Entanglement by measurement}
\label{entanglement}

Entanglement is generated naturally when physical systems interact coherently. For example two electrons trapped at nearby sites can entangle with one another through their mutual Heisenberg interaction. This is the basis of many schemes for QIP, especially in the solid state. However, there is another route to entanglement: rather than allowing the physical qubits to interact directly, instead they can be kept at fixed, well separated sites and simply subjected to a measurement.

Typically one would employ the following idea. The basic unit is an optically active entity which we call the {\em matter qubit}: it could be  an atom, or an atom-like structure such as a quantum dot or an optically active crystal defect. Following laser stimulation the matter qubit may emit a photon; its internal state then depends on the presence/absence of such a photon, or on photon polarisation, etc. To generate entanglement between two matter qubits we would initialize them and then subject them both to laser excitation simultaneously. We would monitor their emissions in such a way that we infer characteristics of their mutual state {\em without} learning the source of any given photon. The essential `trick' in most schemes, including the original 1999 proposal of Cabrillo {\em et al}~\cite{Cabrillo99}, is to use a beam splitter in the apparatus so that any detected photon {\em could have} originated from either source qubit. This technique is referred to as {\em path erasure}.

\begin{figure}[!h]
\centering
\includegraphics[width = 0.8 \columnwidth]{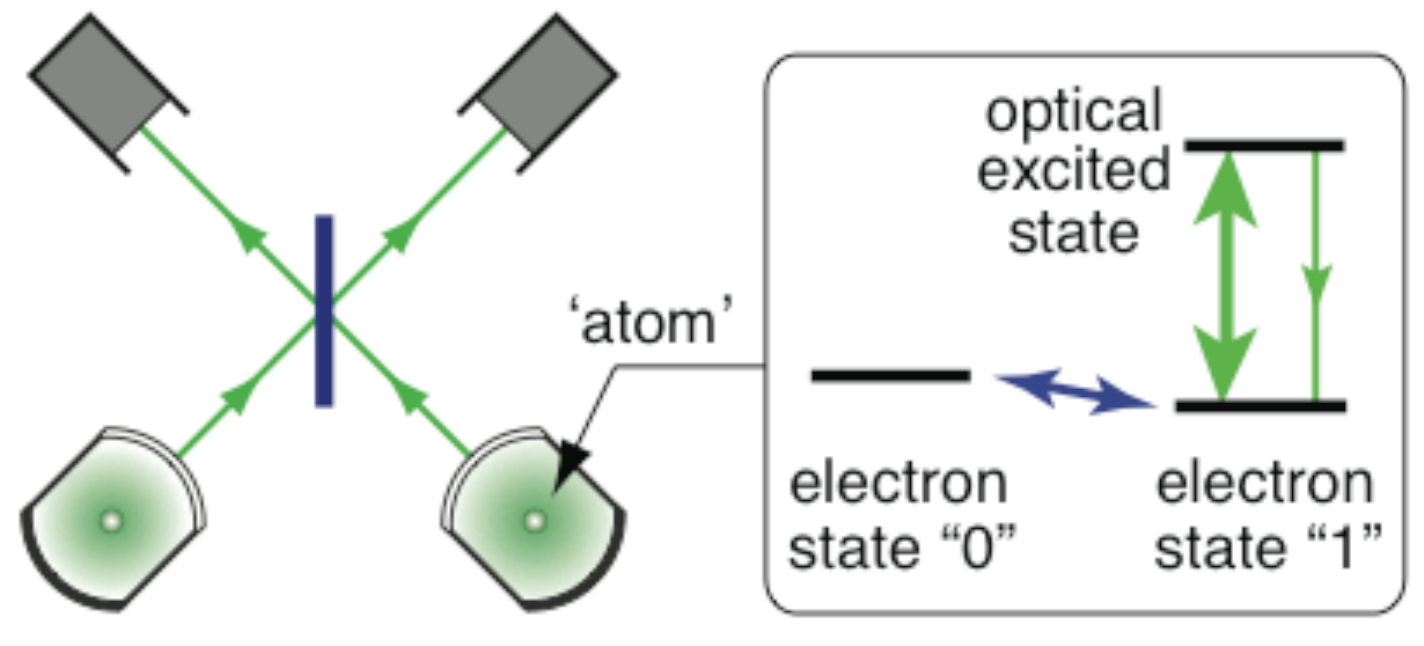}
\figcaption{An apparatus capable of creating entanglement through a measurement.}
\label{erasure_fig}
\end{figure}

Consider the simple schematic diagram shown in Fig.~\ref{erasure_fig}. Suppose that each of the matter qubits A and B has the ``L'' level structure shown on the right. For a moment let us also imagine that the apparatus is ideal, in that no photon will ever be lost from the system. Any photon emitted by the matter qubits will eventually reach the detectors and register a `click'. Moreover we will assume that the detectors can count the number of incident photons (distinguishing zero, one and two). Let us perform a thought experiment with the following steps: (1) We initialize each matter qubit to the state $\ket{+}\equiv{1\over \sqrt{2}}(\ket{0}+\ket{1})$. (2) We subject both matter qubits to a laser pulse, performing a $\pi$ rotation on the transition $\ket{1}\rightarrow\ket{e}$. Now, at this stage the state of each qubit is ${1\over\sqrt{2}}(\ket{0}+\ket{e})$, so that the state of the entire system is
\begin{equation}
{1\over{2}}(\ket{00}+\ket{0e}+\ket{e0}+\ket{ee})
\end{equation}
where we have simply multiplied out the states of the two separate systems A and B, and have organized the qubit values in each `ket' as $\ket{AB}$ (keeping track of the ordering of the labels may appear inconsequential at present, but our reason for doing so will become clear in the next section). We can now think about the future evolution of each term in this superposition. Firstly, we see that the term $\ket{00}$ has no excitation, and therefore no photon will emerge from the matter qubits and the detectors will never `click'. We also note that state $\ket{ee}$ will eventually produce two photons, as the two excited matter systems relax to state $\ket{11}$. These photons will eventually be seen by the detectors (since we have no loss, and the capability to count). In fact a final detector reading of zero indicates that the matter qubits are in state $\ket{00}$, while a final detector reading of two implies that the matter qubits are in state $\ket{11}$. Neither of these are interesting -- we could have prepared those states directly in the first place! However, what of the other two terms? States $\ket{0e}$ and $\ket{e0}$ will both produce one photon, and so a detector reading of {\em one} need not differentiate between them. So what is the final state of the matter qubits in the event that the detectors register a total of {\em one} photon? To answer this we have to track the evolution of the photon. For simplicity we imagine that at a given moment the photon has been emitted from the matter qubit (whether it is the left or right system) but has not yet passed the beam splitter. At this time the state has evolved as follows:
\begin{equation}
\left(\ket{0e}+\ket{e0}\right)\ket{\rm vac}\rightarrow\left(\ket{01}a_R^\dagger+\ket{10}a_L^\dagger\right)\ket{{\rm vac}} \label{vacEqn}
\end{equation}
where $\ket{\rm vac}$ is the vacuum state of the electromagnetic field. We introduce the photon creation operator $a^\dagger_L$ to represent a photon in the lower left channel in Fig.~\ref{erasure_fig}, and similarly $a^\dagger_R$ in the right channel. For simplicity we will neglect to write the $\ket{\rm vac}$ explicitly in subsequent expressions; it should be understood to be present on the right side of the expression just as in (\ref{vacEqn}). Now if the photons in the two $a$ channels are formally indistiguishable (in all senses except for the channel they occupy!), and they are incident on the beamsplitter in such a way as to map onto the same two output modes $b_L^\dagger$ and $b_R^\dagger$, then $a_L^\dagger$ and $a_R^\dagger$ are transformed by the beamsplitter as as follows:
\begin{equation}
a_L^\dagger\rightarrow {1\over\sqrt{2}}(ib_L^\dagger+b_R^\dagger)\ \ \ \ \
a_R^\dagger\rightarrow {1\over\sqrt{2}}(b_L^\dagger+ib_R^\dagger) \label{eqnthree}
\end{equation}
where the phase $i$ is associated with reflection rather than transmission. Using this transformation we can write the state of the system once the photon has passed the splitter:
\begin{eqnarray}
&\ &\ket{01}(b_L^\dagger+ib_R^\dagger)+\ket{10}(ib_L^\dagger+b_R^\dagger)\\
&=&(\ket{01}+i\ket{10})b_L^\dagger+(i\ket{01}+\ket{10})b_R^\dagger \label{eqnfive}
\end{eqnarray}
where we have simply collected terms involving $b_L^\dagger$ and $b_R^\dagger$ and neglected the overall $1\over\sqrt {2}$. Now the action of the detectors is precisely to record either a photon corresponding to  $b_L^\dagger$, or one corresponding to $b_R^\dagger$. Suppose that in fact the left detector clicks -- then we have projected the matter qubits into the state ${1\over \sqrt{2}}(\ket{01}+i\ket{10})$, inserting the correct normalisation. Thus, if we see this measurement outcome then we have projected the matter qubits onto an entangled state -- in fact, a maximally entangled state. Similarly a single click in the right detector heralds the matter qubit state ${1\over \sqrt{2}}(i\ket{01}+\ket{10})$. Notice that it is essential to know {\em which} detector clicked, because the two possible resulting states have different phases, and in fact if we write down the {\em mixed state} corresponding to uncertainty about which detector clicked, then this mixed state has no entanglement at all.

\subsubsection{Parity projection}
We could not achieve much of significance if we were limited to creating two-qubit entangled states using this method, but fortunately it turns out that repeating the same process allows us to create multi-qubit entangled states. To see this imagine that the left hand detector in the previous section has clicked, so that qubits A and B are entangled in state $\ket{01}+i\ket{10}$, and that we initialize a third matter qubit `C' into state $\ket{+}$. The total three-qubit state is now
\begin{equation}
\ket{010}+\ket{011}+i\ket{100}+i\ket{101}
\label{threequbitspre}
\end{equation}
where each three qubit `ket' is now ordered $\ket{ABC}$. We may now subject qubits B and C to a laser pulse inducing $\ket{1}\rightarrow\ket{e}$, and allow the system to evolve using a similar path erasure scheme as described previously. Once again we will have `failure' outcomes corresponding to seeing zero or two photons, this time with the added penalty of destroying the existing entanglement, but in the event that we see exactly one photon then qubits B and C will have become entangled. The final state that will result in only one detector click is
\begin{eqnarray}
&\ &i\ket{101}(b_L^\dagger+ib_R^\dagger)+\ket{010}(ib_L^\dagger+b_R^\dagger)\\
&=&i(\ket{101}+\ket{010})b_L^\dagger+(\ket{010}-\ket{101})b_R^\dagger.
\end{eqnarray}
By recording which detector clicks, we now obtain a three-qubit maximally entangled state. This process of adding a matter qubit to an existing entangled state can be repeated to grow an arbitrarily large entangled state (although as we will see shortly, there are better ways to proceed than to add one qubit at a time). We can express the process of entanglement building as subjecting the two target qubits to a {\em parity projection} leaving them in the odd parity subspace -- i.e. we have excluded the even parity states $\ket{00}$ and $\ket{11}$ from the $\ket{BC}$ subsystem's superposition. Formally, we have applied the operator $\hat{P}_{BC}=\ket{10}\bra{10}+p\ket{01}\bra{01}$ to the separable state in equation \ref{threequbitspre}, where $p$ is either $i$ or $-i$ depending on which detector clicked. Now this operation is ideal for building up a particular kind of multi-qubit state called a {\em graph state} (of which the {\em cluster state} is a special case). Graph states are a resource that permits quantum computing, as we discuss below.

\subsubsection{Photon loss}
As we move from considering an idealized apparatus to a realistic system, the principal issue to address is photon loss. Any photon lost from the apparatus in the scheme described above will leave the matter qubits in an uncertain state. We must of course include detector inefficiency as one form of `loss'. One solution is {\em weak excitation}: in this type of approach we keep the rate of photon generation very low, so that a single detector click is much more likely to have resulted from a single photon being emitted than from two being emitted and one lost ~\cite{Cabrillo99,BKPV01a,BPH01a}. Of course this assumption is imperfect; the fidelity of the final state can only be improved by further reducing the rate of entanglement generation and so there will always be a finite error. Nevertheless, the approach is attractive in its simplicity and was the method used by Chou {\em et al} to entangle remote atomic ensembles in ref.~\cite{Chou05}. A more sophisticated alternative would be to adopt a so-called {\em two-photon} scheme: Here both matter qubits emit a photon, and successful entanglement involves detecting them both~\cite{DK01a,FZLGX01a,SI01a,BK01a,LBK01a,BES01a}. We can rule out the possibility that a photon was lost (since we see one from each source). These schemes do not necessarily require both photons to be present at the same time; they may be emitted in two successive `rounds' as in~\cite{BK01a} for example. Typically in these approaches, success is heralded by detector `clicks' which correspond to single photons impinging on two different detectors, or on a given detector at two distinct times. Consequently there is no need for the detectors to be capable of counting the number of incident photons. Moreover, two-photon schemes are typically {\em interferometrically stable}, meaning that the overall protocol is insensitive to path length variations.  The price to pay for these advantages is that entanglement only occurs when both emitted photons are detected so the success rate falls quadratically with the probability that a given photon is retained. However, even this problem can be mitigated if one can exploit additional complexity within each node (e.g., other eigenstates of the nanostructure)~\cite{CampbellAndBenjamin08}.

The majority of measurement-induced-entanglement schemes fall into these categories, although there are several other approaches -- an example is the idea of scattering a single photon (or stream of such photons) from two physical qubits prior to its measurement~\cite{Protsenko2002,Childress2005}.

It is important to distinguish clearly between the {\em success rate} of an entanglement operation, and the {\em fidelity} of that operation. Generally any given protocol will be able to recognize that certain kinds of error have occurred; in particular, all practical protocols must distinguish when a photon has been lost. When we know that an error has occurred we can designate the operation as a failure and reset our qubits. Thus photon loss affects the success rate but need not affect the fidelity of the successful entanglement operations, when they occur. However, generally there will be other errors which a given protocol cannot detect (or be immune to). One commonly vulnerability is that of {\em dark counts}, where a `click' occurs though no photon was emitted from the matter qubits: on seeing such a count, we may wrongly conclude that we have created entanglement. Experimental defects of this kind will directly impact the fidelity of the entanglement operations, and therefore it is essential to minimize them.

One variant of the two-photon approach was used by Moehring et al in ref.~\cite{Moehring07} to entangle two $^{171}$Yb$^+$ ions at a success rate of about thirty per billion attempts. This low success rate should improve by orders of magnitude with the use of improved apparatus, and in particular the use of cavities to direct emitted photons into the optical apparatus. 
Nevertheless, the low success rate achieved so far does highlight the question of how one can perform a complex computation when the enabling operation, i.e. entanglement generation, is prone to fail. There are essentially two classes of solution: one can adopt a smart growth strategy which allows the construction of large states despite high failure probabilities on each individual operation, or one can use a {\em brokering} protocol which requires a second physical qubit at each location. These two approaches will be addressed in the next section, but first we shall introduce the concepts behind the use of pre-prepared entangled states as a resource for quantum computation.

\subsection{MBQC and graph states}

In 2001, Briegel and Raussendorf reported a radical new approach to quantum computing based on creating a large entangled state and subsequently consuming it by a series of measurements~\cite{RausBriegel01}. They were able to show that this approach permits one to accomplish the same tasks that can be performed by the earlier {\em circuit model} paradigm. Their new paradigm for quantum computation had clear practical potential in the context of optical lattice atom traps, where it may be easier to create many-qubit entangled states than to entangle specific pairs of atoms. But the applicability of the idea is very wide. In particular, it offers a very natural way to proceed when one wishes to exploit {\em measurement induced} entanglement.

\begin{figure}[!h]
\centering
\includegraphics[width =  \columnwidth]{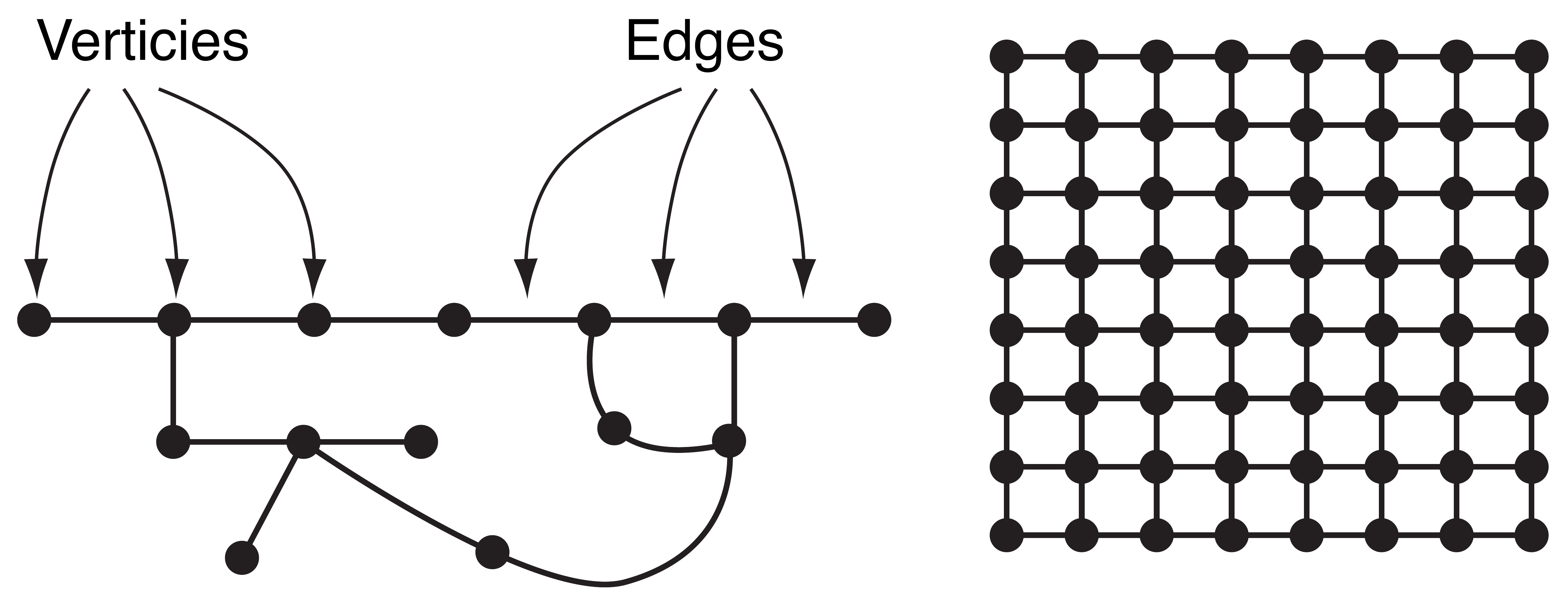}
\caption{Left: a graph state. Right: a cluster state~\cite{RRandBriegelFirstCC}. }
\label{graphExample}
\end{figure}

The kind of state that Briegel and Raussendorf considered can be represented by a diagram that mathematicians call a {\em graph}: A set of points or {\em vertices} connected by lines or {\em edges}. See Fig.~\ref{graphExample} for an example. Graphs are a helpful construction in understanding various aspects of quantum information, for example error-correcting codes~\cite{Schlingemann2001}. Here we wish to associate a quantum state with a graph to obtain what is called a {\em graph state}~\cite{RBB01a,Hein2004}. The most straightforward way to understand the nature of this state is to use the so-called constructive definition: For each node in the graph, prepare a qubit in state $\ket{+}$, and for each edge connecting two nodes in the graph, perform a control-phase gate between the corresponding qubits. Now the control-phase operation is simply the unitary that puts a phase of $-1$ on the $\ket{11}$ component of the system. We shall see presently that this can be replaced with the parity projection that we met earlier. Figure~\ref{graphExample} also shows a {\em cluster state}: a graph state with a regular square lattice, such as the 2D square array.

The simplest non-trivial graph state is just two nodes joined by an edge. The state represented by this graph is
\[
\ket{G}=\frac{1}{2}\left( \ket{00}+ \ket{01}+ \ket{10} - \ket{11}\right)
\]
Note that we only need a single-qubit rotation to transform between this state and the two-qubit entangled states we found resulted from measurement in our beam-splitter device. Any state that can be turned into a graph state just by single qubit gates (unitary gates, not measurements) is said to be {\em local unitary equivalent}, or {\em LU} equivalent, to the graph state~\cite{MvDN05}.

Before reviewing the utility of graph states, let us extend the pictorial notation of figure \ref{graphExample} in a way that will be helpful. We will draw an open circle to represent a physical qubit that is connected to the rest of the graph state by the usual control-phase gate, but which was prepared in some general qubit state $\ket{\psi}=\alpha\ket{0}+\beta\ket{1}$ instead of the standard $\ket{+}$ initialisation. Then of course the total state will not be a graph state, because state $\ket{\psi}$ is not part of the proper definition of a graph state.

\begin{figure}
\centering
\includegraphics[width = 0.95 \columnwidth]{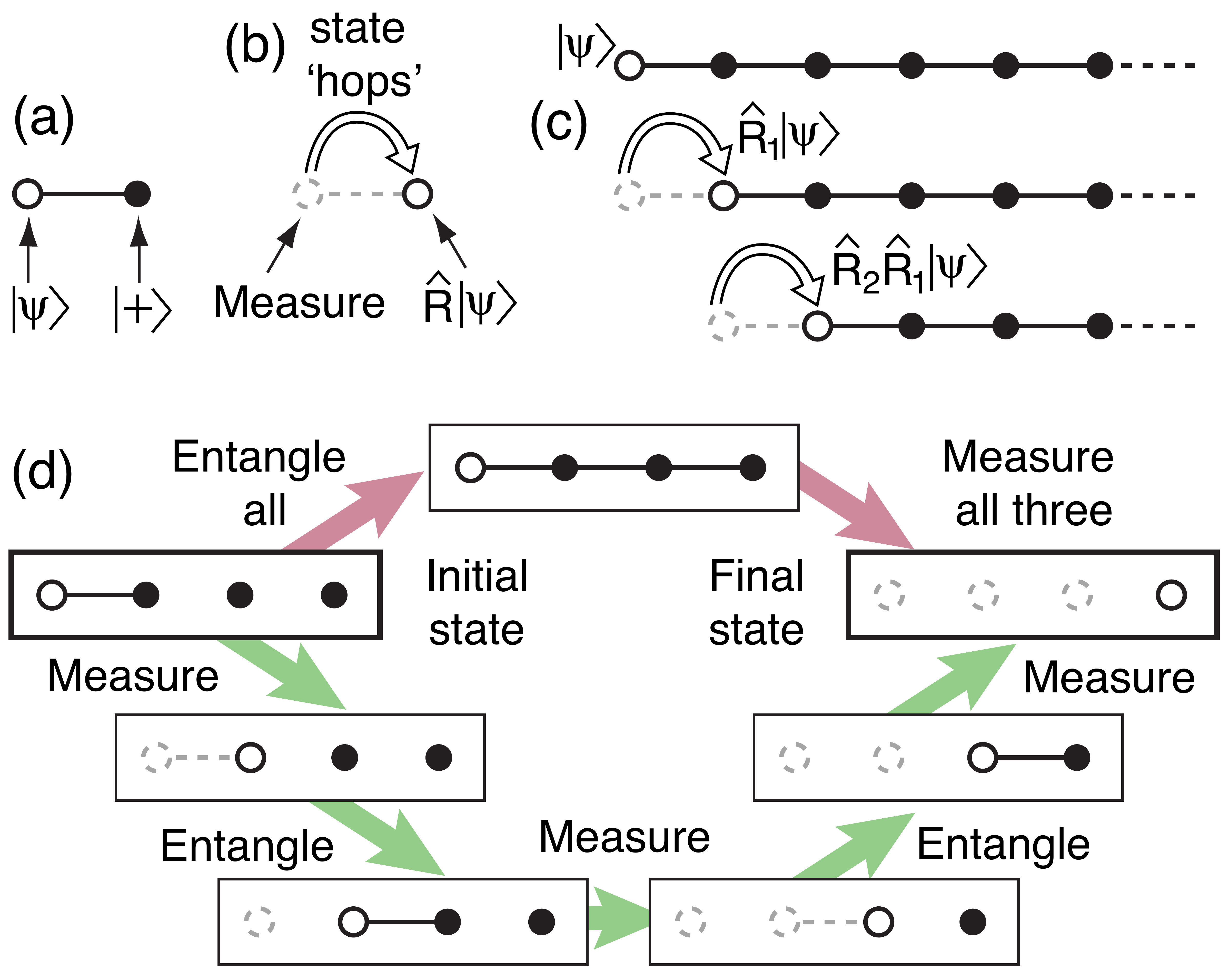}
\caption{(a) A diagram corresponding to a qubit in a general state $\ket{\psi}$ (open circle) entangled with a qubit in the standard $\ket{+}$ state (filled circle). (b) A simple `hop' of one step, resulting from measuring the left-hand qubit. (c) An extended linear graph state can act as a wire, in that a state will hop along it as we measure out qubits. (d) An illustration of why the process of (c) must work: Suppose that, from the initial state, we entangle all the remaining qubits and then measure them out, as shown by the red arrows. In fact the final state must be the same {\em as if} we had performed the alternative time ordering of operations shown by the green arrows. This follows from the fact that the measurement operations commute with entanglement operations on $3^{rd}$ party qubits -- being local to different qubits, they operate on different subspaces of the system's full Hilbert space.}
\label{linearGraph}
\end{figure}

Let us consider just one such general qubit, connected by an `edge' to one regular qubit $\ket{+}$, see Fig.~\ref{linearGraph}. Regarding this diagram as a prescription, it is telling us to (a) take two qubits, (b) put one in some state $\ket{\psi}$ and one in state $\ket{+}$, and then (c) do a control-phase gate between them. Then we have
\begin{equation}
\ket{G'}=\frac{1}{\sqrt 2}\left( \alpha\ket{00}+ \alpha\ket{01}+\beta\ket{10}-\beta\ket{11} \right). \label{qubit-to-plus}
\end{equation}

Now let's measure out the first qubit, the one that was prepared in state $\ket{\psi}$. We measure it in the $x$-basis, so that the outcomes can be  $\ket{+}$ and $\ket{-}$. The two possible states of the remaining qubit are found from $\braket{+}{G'}$ or $\braket{-}{G'}$ with renormalisation, i.e.
\begin{eqnarray}
\frac{\alpha+\beta}{\sqrt 2}\ket{0}+\frac{\alpha-\beta}{\sqrt 2}\ket{1}\ \ \ \ {\rm for\ outcome}\ \ket{+}\\
\frac{\alpha-\beta}{\sqrt 2}\ket{0}+\frac{\alpha+\beta}{\sqrt 2}\ket{1}\ \ \ \ {\rm for\ outcome}\ \ket{-}
\label{hopEqn}
\end{eqnarray}
We see that the state $\ket{\psi}$ has rotated, and hopped or teleported from one physical qubit to the other (Fig.~\ref{linearGraph}b). This connection between the idea of quantum teleportation and graph states is a deep one; indeed teleportation was exploited to achieve computation in the linear-optical scheme of Knill, Laflamme and Milburn (KLM)~\cite{klmPaper} which was proposed around the same time that Ref.~\cite{RausBriegel01} was published. The former approach can be seen as an instance of the latter~\cite{sandu}, although typically one thinks of a different set of allowed primitive operations in the two paradigms.

The same `hopping' works in other cases -- we can use measurements to drive a state from one end of a linear graph state to the other as in Fig.~\ref{linearGraph}(c).  There is an elegant way to verify that this works, using the time-ordering argument depicted as a flow diagram in Fig.~\ref{linearGraph}(d). This diagram illustrates that when we measure out qubits from our complete graph state, the same final state occurs {\em as would occur} if we simply entangle each qubit with its successor and immediately measure it. And since we already know the effect of measuring an entangled pair, consequently it must work completely!
This argument is an instance of a useful general rule: if we want, we can build a graph state at the same time as we are measuring it. The behaviour of the whole system will be just {\em as if} we had built the whole thing first, {\em provided} that whenever we measure a qubit, its entanglement `edges' are already in place to reach its nearest neighbors~\cite{RBB01a}.

Thus a linear graph state can act as a wire for conducting an unknown quantum state -- one can drive the state along by making measurements. The state will get rotated each time it `hops', but as long as we record all the measurement outcomes we can keep track of the cumulative rotation so that we will always be able to `fix' the state when it gets to the other end. The seminal paper of Raussendorf and Briegel~\cite{RausBriegel01} contained a crucial further observation: Suppose we try our same trick of measuring a qubit to make it `hop' along a linear graph state, but this time we don't measure it in the $x$-basis, but instead in some more general basis for which the states $\ket{A}=\frac{1}{\sqrt 2}\left( \ket{0}+e^{i\phi}\ket{1} \right)$ and  $\ket{B}=\frac{1}{\sqrt 2}\left( \ket{0}-e^{i\phi}\ket{1} \right)$
are the possible measurement outcomes. By evaluating $\braket{A}{G'}$ we find that after measurement outcome $\ket{A}$ we are left with
\[
\frac{(\alpha+e^{-i\phi}\beta)\ket{0}+  (\alpha-e^{-i\phi}\beta)\ket{1}}{\sqrt 2}
\]
after normalisation. This is similar to before, the state hops and rotates. But this time it rotates not simply by a Hadamard alone but also another rotation ${\hat U}_z(\phi)$ which incorporates our {\em chosen} parameter $\phi$. And this is interesting, because it shows that we can adjust the rotation by our choice of measurement direction. In fact, it turns out that we can string together three successive measurements in a linear graph to create {\em any} single qubit rotation we wish (up to the usual unwanted cumulative rotation, which we need only track by recording measurement outcomes).

Remembering that all we need for universal QIP is (a) the capacity to perform single qubit gates plus (b) an entangling gate, it is now natural to ask ``can we do an entangling gate, too?"

\subsubsection{Two-dimensional graph states.}

\begin{figure}[h]
\centering
\includegraphics[width = 0.95 \columnwidth]{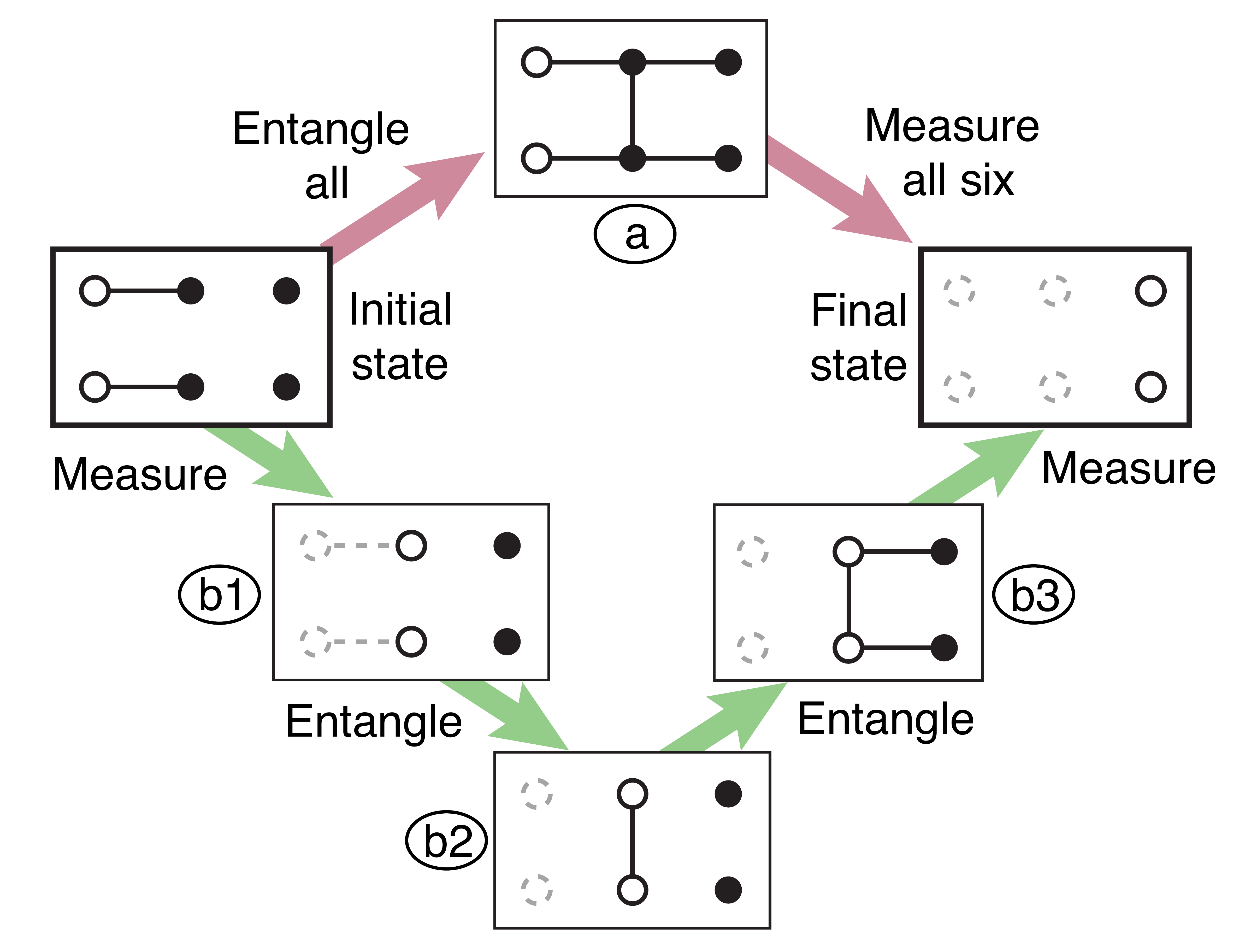}
\caption{Verifying that the 2D graph state formed from connecting two linear graph states can implement an entangling gate.}
\label{twoQubitCircuit}
\end{figure}

We have the idea of a linear graph state as a kind of wire. It is natural to consider two such wires, and then putting a bridge between them. The graph state that we would try would be as shown in Fig.~\ref{twoQubitCircuit}a.

In fact this will indeed suffice to entangle two unknown states. Once again, we can just use our rule about building a graph state as we go: we are allowed to measure a node as soon as all the `edges' out of that particular node are complete. So measuring all the qubits in the state (a) of Fig.~\ref{twoQubitCircuit} must be equivalent to the step-by-step process shown by the green arrows in the lower half of the Figure. We see that the action is like that of two parallel wires, except that at point (b2) the qubits stored on a particular pair of nodes get a phase gate between them. This final observation allows us to see the full power of graph states: if we can make a graph state with the right topology then we can do any quantum algorithm we want, just by making single qubit measurements.

\begin{figure}[h]
\centering
\includegraphics[width = 0.9 \columnwidth]{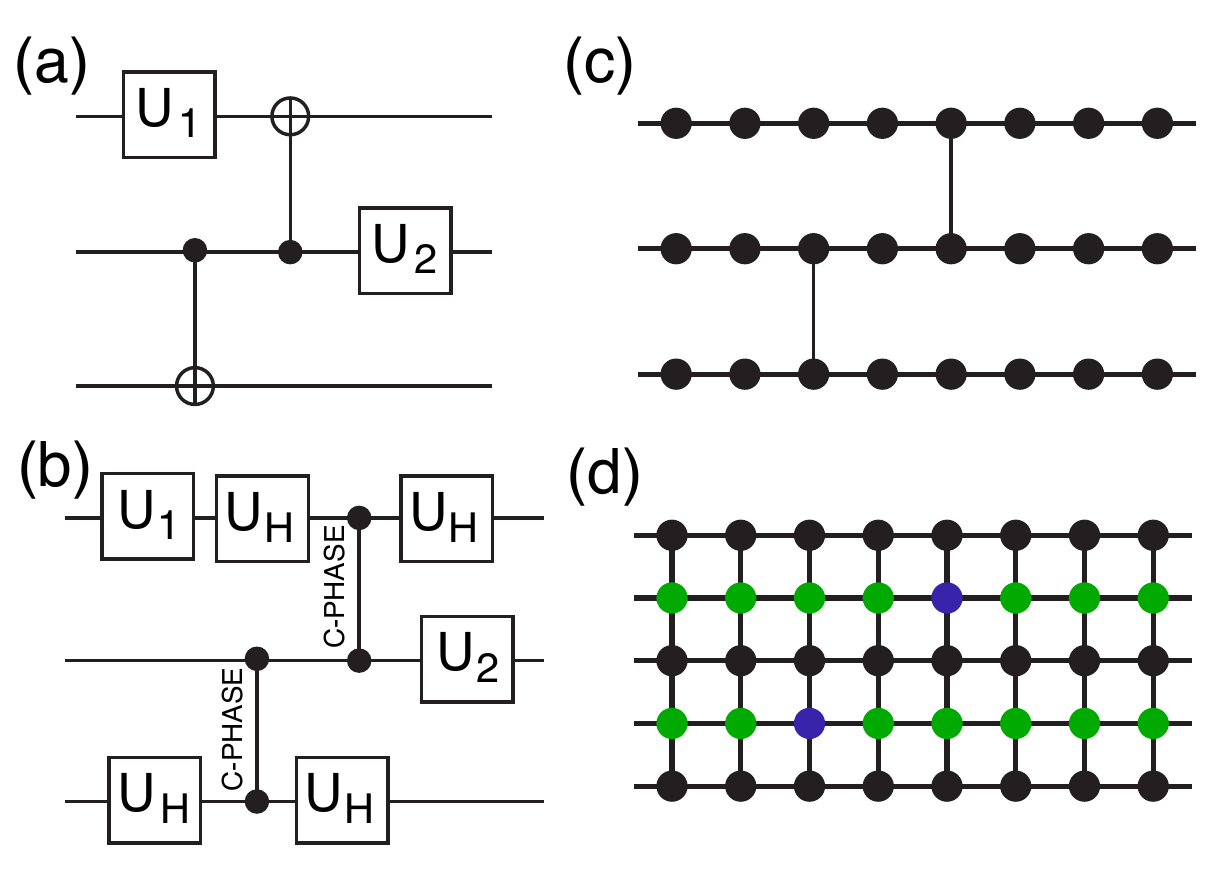}
\caption{Finding a graph state that contains all the entanglement needed for a particular quantum algorithm. Take some desired quantum circuit (a) and translate all the two-qubit gates into control-phase gates (b). Then write down a graph state with the same topology, allowing three vertices in a linear chain for each arbitrary single qubit gate. Also note that one could obtain the same  graph state from a regular 2D cluster state (d), simply by measuring out unwanted qubits in either the $z$ basis or the $y$ basis (green and orange respectively). }\label{universalResource}
\end{figure}

If there is a particular algorithm that we want to perform, then we should be able to draw it as a quantum circuit. Then it's easy to work out at least one particular graph state that is capable of implementing that circuit operation, as shown in Fig.~\ref{universalResource}. One further interesting point is that we can remove any qubit from within a graph state simply by measuring it in the $z$-basis. If the outcome is $\ket{1}$, we apply a further `fix': subject each neighbour to a single qubit phase gates $\sigma_Z$. The result is a new graph state with the node corresponding to the measured qubit removed. Similarly, measuring in the $y$ basis will remove a qubit {\em and} connect its neighbours. Using this principle we can see that a (sufficiently large) cluster state is a universal resource for any algorithm: we simply prune out unwanted qubits until our cluster state becomes the correct graph state for a given task (see Fig.~\ref{universalResource}d).

\subsubsection{Growing graph states} 
\label{growing}
We have seen that a graph state with a suitable topology can allow one to perform a quantum algorithm simply by making measurements. But we have not explicitly shown how to create such a graph state; the constructive definition of a graph state is expressed in terms of a control-phase gate, whereas most schemes for measurement induced entanglement produce a {\em parity projection} (with at least one notable exception~\cite{LBK01a}). Moreover, all realistic schemes have a high probability of failing any given entanglement operation. In this section we note that even a probabilistic parity operation can indeed suffice to efficiently grow graph states.

\begin{figure}[h]
\centering
\includegraphics[width = 1 \columnwidth]{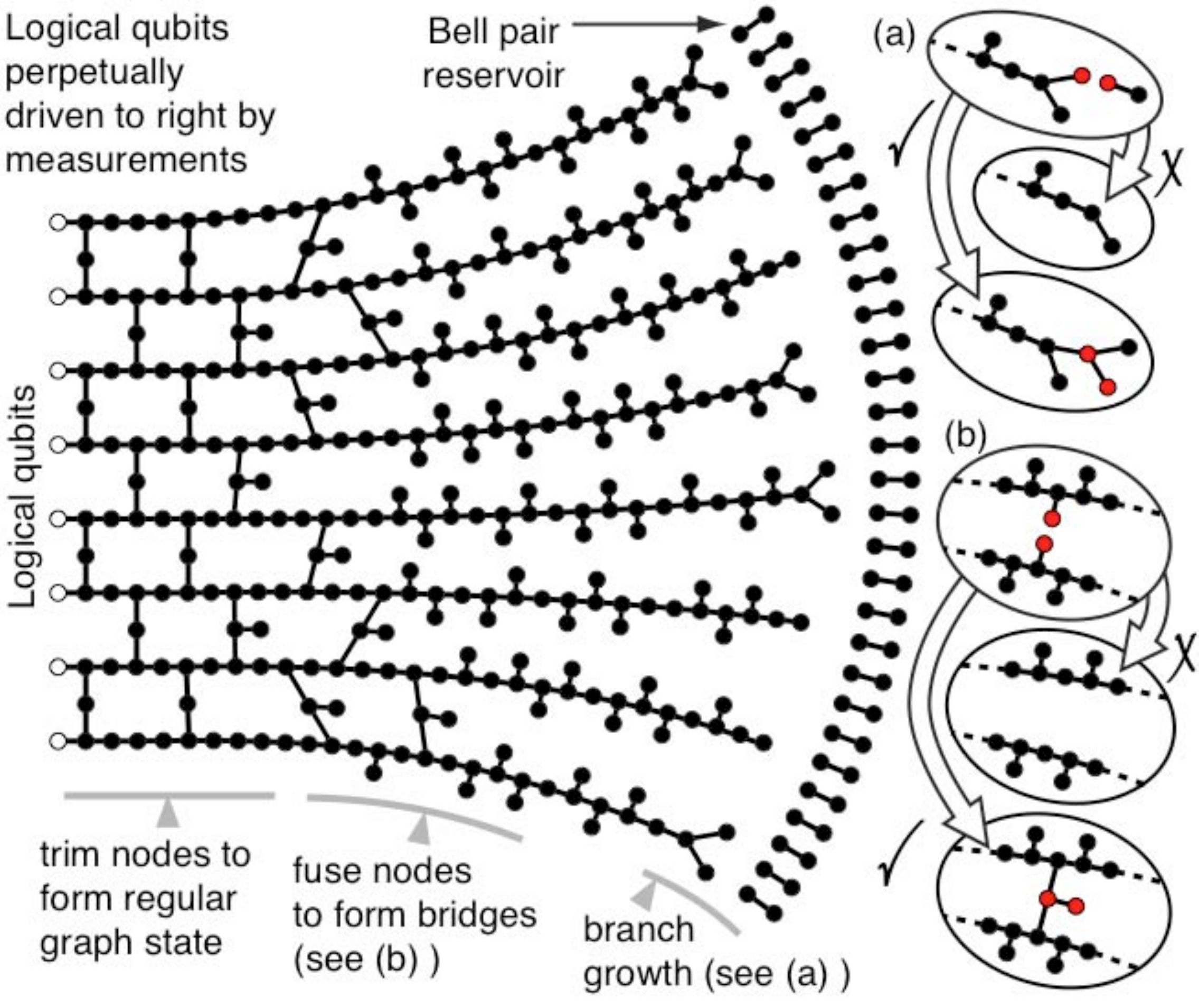}
\caption{A simple example of how one might `grow' a graph state `just in time' for it to be consumed, as described in the text. The regular graph state structure on the left is a universal resource, like the cluster states seen earlier. The insets show the effect of successful and unsuccessful parity projections on qubits (red) that have prior entanglement \cite{Campbell08}.}\label{bramble}
\end{figure}

Figure~\ref{bramble} shows one approach. The main graph state consists of `branches' which grow when successful parity projections add Bell pairs to the `tips'; a failure results in the branch shortening, however since success adds two qubits whereas failure only removes one, we will have average growth provided the probability of success $p$ exceeds $1/3$. We make the branches long enough to absorb an unlucky string of failures. More sophisticated strategies can handle arbitrarily low $p$~\cite{BK,BKcom,DandRplus,Gross06} although in practice the decoherence time of the system will become an issue. Conversely, very rapid graph state growth can occur when $p$ is above a percolation threshold~\cite{perc1,perc2}. There has also been work on graph state synthesis when there are experimental imperfections such as systematic asymmetries in the apparatus~\cite{earl1,earl2}.

\begin{figure}[h]
\centering
\includegraphics[width = 1 \columnwidth]{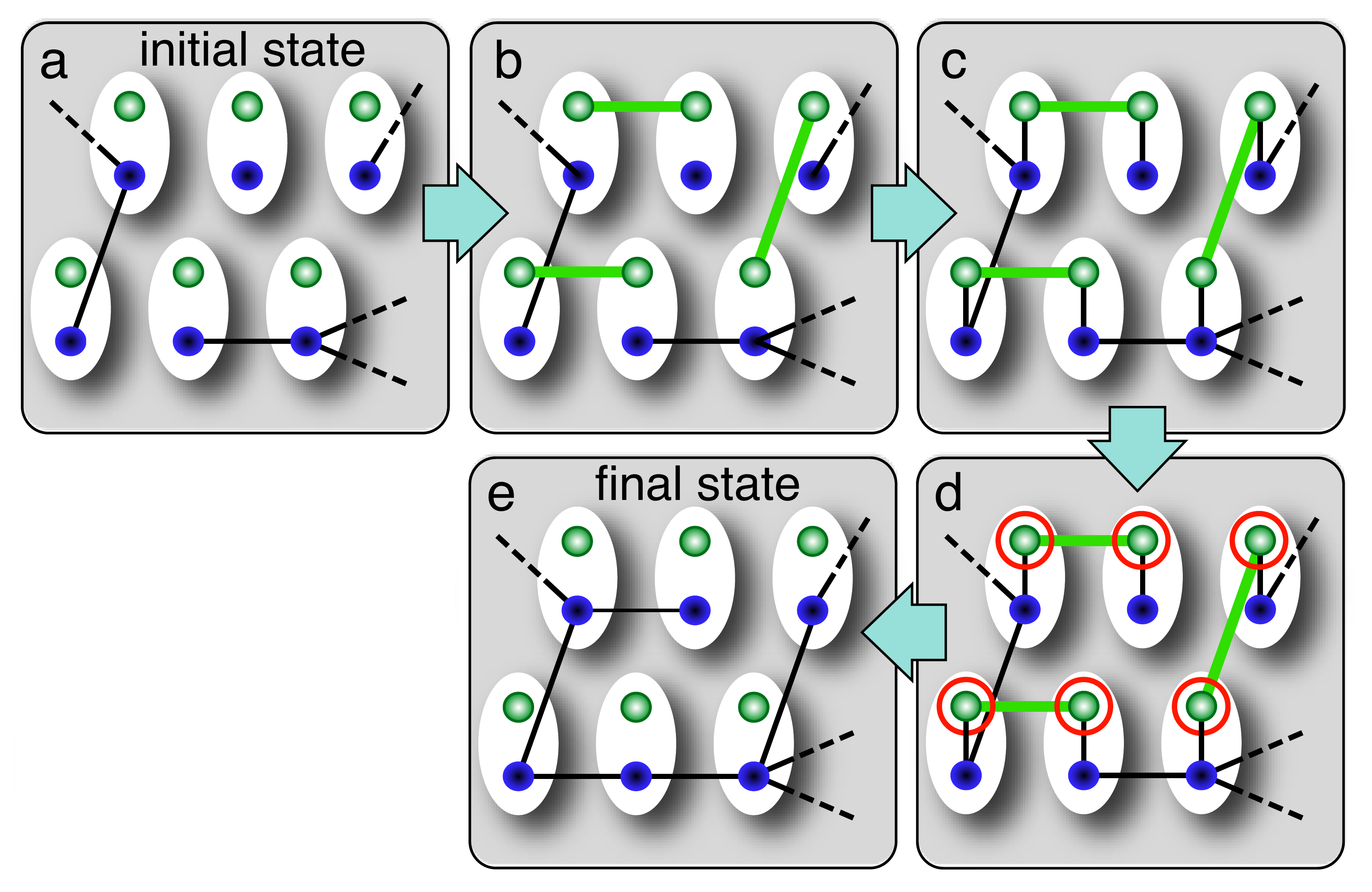}
\caption{Creating a graph state through brokering \cite{brokerPaper}. Each physical node has two qubits, the optically active broker (green) and the passive client (blue). The role of the client qubits is to hold the large scale graph state. Here we wish to reach final state (e) from initial state (a) without risking the loss of the existing entanglement. The broker qubits are projected into entangled pairs (b) so that the desired new entanglement `edges' exist in the broker space. Now the brokers are entangled with their client by deterministic local operations (c), and finally (d) the brokers are measured out in the $y$-basis in order to project their entanglement onto the clients. }
\label{brokering}
\end{figure}

When $p$ is very low, for example due to high photon loss, then it becomes practically essential to find a way of preventing failures from damaging the nascent graph state. The solution is {\em brokering}~\cite{brokerPaper}. This requires that our basic physical system has at least two qubits within it, for example an electron spin (with associated optical transition as in Fig.~\ref{erasure_fig}) and a nuclear spin. We assume that the two qubits in each location can controllably interact, so that within each elementary node of our device we are free to perform one- and two-qubit operations deterministically and with high fidelity. Given this level of resource, we can use the optically active qubits to achieve entanglement between nodes (a process that may fail many times before succeeding) and then transfer that entanglement to the second qubits which are actually storing the large scale graph state. Thus the optically active qubits act as entanglement `brokers', insulating their partner `client' qubits from the many failures they may encounter before successfully becoming entangled. A simplified form of this process is depicted in Fig.~\ref{brokering}. Subsequent work has shown that two qubits also suffice for certain kinds of entanglement distillation, including procedures to combat phase noise~\cite{earlsOwn} and corruption due to photon loss~\cite{CampbellAndBenjamin08}. If the physical system is such that more than two qubits exist at each site, then there are potentially further advantages such as general entanglement distillation~\cite{BriegelDistil,LukinDistil}.

%%%%%%%%%%%%%%%%%%%
%% BENJAMINS

\section{Solid state architectures for MBQC: General considerations}
\label{spins}

In many ways, solid state systems offer an ideal realization of  
measurement based quantum computing. Whereas free ions need a complex  
array of electromagnets and ultra-low temperatures in order to be  
positioned for accurate control, solid state qubits are stationary and  
can be individually addressed even at room temperature. Nano- 
fabrication techniques also allow for the creation of arrays of  
qubits, or the isolation of a single qubit, architectures that might  
both be used in a measurement-based device. The suitability of  
spatially remote qubits for measurement-based protocols means that  
many of the aspects of solid state qubits that present major headaches  
for circuit-based schemes, such as low qubit yield and the need to  
fully characterise each qubit, can be accommodated.

Some drawbacks remain, however. The most serious of these is that  
there are many different uncontrolled interactions between the qubit  
and  the various degrees of freedom that exist in its environment.  
These can lead to a number of problems. First, the coupling between  
the qubit spin state and its surroundings cause a leaking of quantum  
information known as decoherence. Second, optical transitions can  
fluctuate in energy on a wide range of time scales as a result of  
noise in the local environment and cause an uncertain phase relation  
between remote centres. Third, a loss of fidelity in single qubit  
rotations which, like decoherence, occurs due to fluctuations in spin  
transition energies resulting from magnetic noise, but presents an  
additional source of errors.
Control of photons is also a challenge. An optical cavity would be  
used to improve the photon collection efficiency but absorption,  
scattering and interface effects in solids can limit their finesse.  
This provides an additional source of inhomogeneity between the qubit  
control modules, which can affect the success of the path erasure  
protocol. Spurious `dark' counts in the photodetectors introduce  
errors by falsely heralding success in entanglement operations.  
However, existing single photon counting technology, particularly for  
wavelengths in the range 400 nm $<  \lambda < 1~\mu$m using silicon  
devices, provides dark count probabilities of order $10^{-8}$ in a typical spontaneous emittion lifetime, whilst  
simultaneously benefiting from the highest detection efficiencies of  
up to 70\%, and so this source of errors can probably be neglected  
relative to the others mentioned above. The aggregate of all these  
errors must be kept within the limits imposed by the overall fault  
tolerance of the scheme, typically a percent or less 
\cite{raussendorf07a,raussendorf07b}.

We next discuss the most important environmental interaction  
mechanisms which may occur for optically measured spin qubits, before  
introducing the physics relevant to optical cavities suitable for  
solid state qubits emitting in the visible or near infrared.
We shall focus in particular on two physical examples, NV centres in  
diamond and III-V semiconductor quantum dots, which will be the main  
topic of section~\ref{qubits}.

\subsection{Decoherence and related effects}

The discussion of decoherence is broken down into three subsections.  
The first two involve spin decoherence, that is to say decoherence in  
the spin basis, resulting from interactions with lattice vibrations  
and with other neighbouring spins. Such decoherence mechanisms are  
important not only to MBQC but to any spin-based quantum computing  
architecture, and have been thoroughly reviewed in recent works by  
Hanson {\em et al.}~\cite{hanson07,hanson08}) so we shall not dwell  
long on them here. The third subsection concerns decoherence in the  
optically excited state $\ket{e}$, which is of particular relevance to  
measurement-based quantum computing architectures.

\subsubsection{Lattice vibrations}

Lattice vibrations (phonons) are not magnetic particles and therefore  
do not directly couple to spin. Rather, they cause small electric  
field fluctuations that would not couple to spin at all, if it were  
not for the {\em spin-orbit interaction}. This links together the spin  
and spatial degrees of freedom such that a changing a spin eigenstate  
will typically also mean changing the orbital wavefunction slightly.  
The coupling originates from the effect of an electron, which is  
charged, moving through the electric field of the surrounding crystal  
environment. This motion bends the orbit of an electron, leading to a  
magnetic moment that can interact with the spin magnetic moment. The  
size of the interaction varies considerably; it can be very small in  
certain materials (e.g. in diamond, spin-orbit interaction is weak  
enough to be neglible for most purposes) - but quite large in others  
(e.g. in GaAs we shall see that it is very important).

There are two different classes of spin decoherence: relaxation and  
dephasing; any error process that occurs within a single qubit Hilbert  
space can be written in terms of these two classes. Relaxation is spin  
depolarization - i.e. spin up states change to spin down, with a  
change in energy, and vice versa -- until a thermal equilibrium is set  
up. Dephasing describes the loss of phase coherence of the spin state  
-- i.e. the relative phase of spin up and spin down components of the  
wavefunction becomes scrambled; this type of decoherence does not  
involve a change in energy and therefore usually occurs on a shorter  
timescale.

The relaxation rate $1/T_1$ can be calculated using Fermi's Golden  
Rule, which tells us that the rate will depend on the electron-phonon  
matrix element that couples the spin up and spin down levels, as well  
as on the density of phonon states. There are two ways in which  
phonons can produce electric field fluctuations, and each has a  
different matrix element. First, deformation of the crystal lattice  
can lead directly to band gap modification, and second the strain  
caused by phonons can give rise to electric fields in polar systems  
through piezoelectric coupling. Plugging all of the factors into  
Fermi's Golden Rule gives a relaxation rate that depends on the Zeeman  
splitting of the spin sublevels, and so on the applied magnetic field  
$B$. The dependence is $B^5$ for piezo-electric coupling and as $B^7$  
for deformation potential coupling. Hence this decoherence mechanism  
is much more important at high fields, where it can reduce the  
relaxation time to less than a tenth of a millisecond~ 
\cite{kroutvar04,amasha08}.

Any  $T_1$ process automatically creates a dephasing, or $T_2$,  
channel. However, on top of this there are additional ways in which $1/ 
T_2$ can become larger, and these contributions give {\it pure  
dephasing}. The $T_1$ and $T_2$ are then related:
\begin{equation}
\frac{1}{T_2} = \frac{1}{2 T_1} + \frac{1}{T_2^{pdp}},
\end{equation}
where $ \frac{1}{T_2^{pdp}}$ is the pure dephasing contribution.  
Interestingly, for electron spins in quantum dots, the pure dephasing  
contribution is predicted to be zero, to lowest order in the spin- 
orbit coupling~\cite{golovach04}.

\subsubsection{Other spins}
\label{decoherencespins}

The other major headache for the designer of spin-based quantum  
computers is the existence of other spins that do not form part of the  
quantum processor. Even though they are not part of the computer  
architecture, spins in the surrounding matrix can nonetheless interact  
with spin qubits and degrade the information contained within them.  
There are two types: electronic and nuclear. Of course, the nuclear  
spin has a much smaller magnetic moment than the electron spin -- but  
each type can play an important role.

Let us first consider nuclear spins. In many candidate solid state  
systems, particular examples being the III-V and II-VI semiconductors,  
almost all the nuclei have a magnetic moment. The dominant interaction  
with an electron is the hyperfine coupling, which occurs when the  
electron wavefunction has a significant amplitude at the position of  
the atomic nucleus. In the solid state, an electron is often quite  
delocalized: for example in a quantum dot it can overlap with tens of  
thousands of nuclei. Such a system of many nuclei and just a single  
electron spin is a highly complex many-body problem, but happily the  
physics can often be described well using a huge simplification: that  
all of the electron-nuclear hyperfine interactions can be thought of  
as a single effective magnetic field, the Overhauser field, acting on  
the electron. The value and direction of this field is not normally  
known in an experiment and the Zeeman splitting of electron spin in an  
applied magnetic field is changed. This change in the field causes an  
unwanted extra phase to accumulate in a spin superposition, and if the  
phase is unknown this is equivalent to dephasing. However, some  
experimental groups have found several ways round this problem. First,  
a `refocusing' pulse sequence can be applied to the electron spin,  
causing it to flip its direction by 180 degrees, and the extra phase  
to unwind. This causes a `spin echo' once the phase has fully  
reversed, and the phase coherence is restored. This kind of  
inhomogeneous dephasing is not then a true decoherence, but it is  
often a limiting factor and the timescale over which is acts, called  
$T_2^\ast$, can be as short as a few nanoseconds~\cite{petta05}.  
Another way around the Overhauser field problem is to control it by  
polarizing the nuclear spin bath; recent impressive measurements have  
shown that this can indeed be achieved by clever manipulation of the  
electron spin qubit itself~\cite{Feng07,reilly08}.

Even with these ingenious tricks, nuclear hyperfine coupling can still  
be a problem, since the nuclei themselves fluctuate, albeit on a  
rather longer timescale. This typically limits the phase coherence of  
spin qubits to times of order microseconds~\cite{petta05}. For this  
reason, it is a very attractive proposition to turn to materials where  
the most common isotopes have zero nuclear spin, for example silicon  
and carbon. If this can not be achieved, substantial improvements in  
$T_2$ may be observed by using electrons with very small wavefunction  
amplitude at the nuclei to reduce the coupling; this would be typical  
of the $p$-symmetry of hole states in III-V semiconductors.

If nuclear spin decoherence can be overcome or eliminated, it can be  
the electron spins associated with impurities that are the main spin  
decoherence channel. The number of defect electron spins in a sample  
is typically many orders of magnitude smaller than the number of  
nuclei. However, this does not make electron spin defects harmless;  
their dipole moment is two thousand times that of a typical nucleus,  
and therefore the dipole-dipole interaction with an electron spin  
qubit can be quite significant. Unlike with nuclear spins, there is  
also the possibility of direct resonant `flip-flop' transitions of a  
defect spin with the qubit electron spin itself. Together, these  
effects can cause dephasing times on the nanosecond scale.  
Fortunately, there are again ways around this. In particular,  
polarization of the defect spins, which for electrons can be achieved  
simply by applying a large magnetic field at low temperature, can  
increase the decoherence time by several orders of magnitude~ 
\cite{takahashi08}.

\subsubsection{Decoherence and energy fluctuations in the optically  
excited state}
\label{excitedstate}
We have so far considered decoherence processes that occur directly on  
the spin qubits. However, in measurement -based schemes the operations  
that generate spin entanglement often involve excitation outside of  
the qubit Hilbert space and then the principal decoherence mechanisms  
are likely to be different.

For example, many ideas rely on photon emission and a higher energy  
state must be excited optically as part of the measurement process. We  
must take account of any new mechanisms that occur whilst such an  
optical excitation is being performed.  To lowest order, such a  
transition would need to be electric dipole allowed - meaning that a  
direct coupling to the electric field fluctuations caused by phonons  
is possible. A detailed calculation of the open system dynamics can be  
done using a density matrix master equation technique~\cite{gauger08},  
and this reveals that the decoherence rate depends on several factors.  
First, the spectral density of the phonon coupling, which  
characterizes the density of states and coupling strength of the  
phonons at a particular energy is crucial, with the relevant energy  
corresponding to the Rabi frequency that characterizes the rate of  
excitation of the higher level. Second, the number of phonons, which  
of course depends on temperature through the Bose-Einstein  
distribution function, affects both phonon absorption and phonon  
emission. Though this is really a $T_1$ type process, it can look like  
a $T_2$ process since the bare energy levels of the qubit system are  
`dressed' by photons under laser excitation. It typically gives a  
dephasing on the sub-nanosecond timescale, though this can be  
increased by using adiabatic methods for eigenstate following~ 
\cite{gauger08,lovett05,calarco03}. The decoherence time of course  
sets an upper limit on the time that a laser can be applied to the  
qubit, if quantum coherence is to be preserved.

A particularly strong phonon interaction can occur in certain  
nanosystems, such as crystal defects, that are accompanied by local  
distortion of the crystal lattice. Such distortion may lead to local  
phonon modes, as opposed to the more commonly discussed bulk modes,  
which have a much larger amplitude at the qubit than their bulk  
cousins, so that coupling is increased. They are also more strongly  
confined, which discretizes their excitation spectrum. It is then more  
profitable to consider these quantized levels to be part of the level  
structure of the qubit~\cite{davies76}, and special measures must be  
taken to avoid populating those levels which have significant phonon  
character.

If the energies of the electric dipole transition energies in two nanostructures
are not equal, then this will introduce an extra phase term into the resulting
entangled spin state. However if the arrival of the photon can be measured with
a timing resolution faster than that in which the phase changes significantly,
this error source can be suppressed~\cite{metz08}. The best timing resolution
for high efficiency single photon detectors is currently of order 50 ps
suggesting that the mismatch should not be any more than a few tens of micro-electron
volts. The additional information recorded using fast detection
is also able to remove the problems of other mismatched parameters, for example
transition dipole strength~\cite{earl1,earl2}.

Electric dipole transition energies can also be affected by randomly  
fluctuating local electric fields caused by the movement of carriers.  
In well designed experiments with high quality samples and minimal  
extraneous excitation, the time scale of such fluctuations can be  
greater than $T_1$ for the optical transition and so do not to affect  
the coherence of emitted photons, as demonstrated by Santori {\em et  
al} in their 2002 report of indistinguishable photons emitted from a  
single quantum dot \cite{Santori02}. However optical transition line  
widths measured in absorption and PLE experiments are often somewhat  
larger than the relaxation rate $1/T_1$~\cite{Seidl05,Batalov08}  
suggesting that over millisecond time scales some `spectral drift' is  
encountered. This must be avoided in order to permit the sustained  
energetic resonance between the optical transitions of individual  
qubits that will be vital to `path erasure' and thus to high fidelity  
entanglement generation.

\subsection{Photon control in the solid state}

The optical cavities depicted in the title figure are not essential  
to the measurement-based entanglement schemes described in section  
\ref{entanglement}, but they can offer the substantial advantage of an  
increased measurement efficiency that would speed up the building of  
entangled states. The role of the cavity is thus to encourage emission into the desired optical mode so that the photon  
can be delivered efficiently to the multiplexer. It can offer an additional benefit of reducing  
the length of time for which the optically excited state is populated, thereby  
limiting the effect of relative dephasing of the two qubits.

There are essentially two ways in which cavities could be employed.  
One would be to create a cavity mode that is resonant with the  
emitting dipole, and make use of the Purcell effect to encourage rapid  
spontaneous emission from the optically excited state into the desired  
mode; this is an established technique in both atomic and some solid  
state systems \cite{Vahala03}. The other is to use the cavity mode as  
a receiving mode for a stimulated Raman adiabatic passage (STIRAP)  
process \cite{Bergmann95}, in which the optically excited state is  
never populated at all and the energy from the coherent excitation  
beam is transferred directly into the output mode. Such processes have  
been demonstrated in atomic systems \cite{Hennrich00}, and could  
potentially be applied in the solid state if sufficient control can be  
exercised on the optical transitions and cavity modes. If realized,  
STIRAP may offer the possibility of eliminating the excited state  
dephasing process completely.

These two modes of operation place different requirements on the  
characteristics of the cavity employed. For enhanced stimulated  
emission, so-called weak coupling is required, in which the cavity  
field leakage rate $\kappa$ is greater than the atom-cavity coupling  
strength $g$, but the atom-cavity relaxation rate, $g^2/\kappa$ should  
be large compared with the spontaneous emission rate in the absence of  
a cavity (usually between 10MHz and 10GHz) so that the photons are  
channeled preferentially into the cavity mode. The important figure of  
merit for the cavity is the Purcell factor $F_P = 3Q\lambda^3/ 
(4\pi^2n^3V)$ where $Q$ is the cavity quality factor, $V$ is the mode  
volume, and $n$ is the refractive index of the material in which the  
cavity mode resides. For STIRAP, so-called strong coupling is  
desirable ($\kappa<<g$), to couple the excited Raman level efficiently  
with the cavity mode. This is a challenging condition to meet in solid  
state microcavities, since $g$ scales with $V^{-1/2}$ and reducing the  
mode volume is best achieved by shortening the cavity round-trip  
propagation distance, which results in a commensurate increase in $ 
\kappa$.

Solid state microcavities operating in the visible and near-infrared  
can offer $Q$ factors up to $10^{10}$, and mode volumes down to  
$~0.01\lambda^3$ but unfortunately no one cavity design combines these  
two extremes \cite{Vahala03}. The highest $Q$'s are  
achieved by whispering gallery modes of transparent microspheres and  
disks or rings. However the accompanying mode volumes are large at $ 
\sim10^3 \lambda^3$ giving $\kappa \sim 10$~MHz and they are better  
suited to strong coupling than to weak coupling applications. Fabry  
Perot cavities such as monolithic micropillars and tunable vacuum- 
based cavities containing solid state emitters can provide $Q \sim  
10^4$ combined with $V \sim 10 \lambda^3$, with the added attraction  
of ease of coupling into the collection optics. Photonic crystal  
cavities offer mode volumes down to $V \sim \lambda^3$ and are  
predicted to provide $Q$ factors up to $\sim 10^6$,\cite{Bayn08} although  
fabrication difficulties have meant that the best achieved to date is   
$\sim 10^5$ using a silicon-on-insulator architecture \cite{Zain08}.  
Finally, cavities based on surface plasmons have potential to offer  
mode volumes down to $~0.01 \lambda^3$, albeit with relatively low $Q$  
factors $\sim 100$ \cite{Wang05}.

Fabricating a monolithic microcavity to provide a mode at a precisely  
prescribed wavelength is immensely challenging, and so if a cavity  
design with a high $Q$ factor is to be employed then it becomes  
important to be able to tune the mode frequency into resonance with  
the optical transition of the qubit. Air-gap Fabry Perots are one  
possible solution to this as they allow the cavity length to be  
adjusted with sub-nm precision using piezoelectric actuators. Fine  
control of the cavity modes and the spatial alignment of the dipoles  
within them is important even if the qubit transitions themselves are  
perfectly at resonance with each other, since path erasure can be  
compromised by temporal inhomogeneity between the photon wavepackets  
resulting from differing values of the enhanced emission rate  
$2g^2/\kappa$ \cite{Campbell08}. Special attention should be paid to  
this issue since loss of path erasure would lead to false `positive'  
photon detection events that would destroy entanglement in an  
unheralded manner.

Finally in this section we make a note regarding the stability  
requirements of the photon interferometer depicted in the title  
figure. In general, it is sufficient that the path lengths from the  
qubits to the interferometer are known to within a small fraction of  
the coherence length of the emitted photon, which in most cases will  
be of order centimetres and therefore easy to achieve. Inequality in  
the lengths themselves is not a problem as it is only necessary that  
the photons arrive at the beam splitter together, which can always be  
achieved by appropriate timing of the excitation pulses to the qubits.  
There is one situation in which short term phase stability is required  
however. In section \ref{entanglement} we noted that the two qubits  
are projected into a different entangled state depending on which  
detector registered a photon. Thus in schemes that require \textit{two  
photons to be detected as a result of different excitation pulses}  
\cite{BK01a,LBK01a}, the phase relationship between the photon  
arriving at the beamsplitter and the emitting qubit must not change  
between the two excitations.  This time may be of order 100 ns or less  
however, and so such schemes may not preclude the use of fibre  
waveguides in which phase stability over longer time scales is poor.

\section{Candidate solid state qubits}
\label{qubits}
\subsection{Nitrogen-vacancy defects in diamond}
\label{NV}
\begin{figure}[h]
\centering
\includegraphics[width = 0.5 \columnwidth]{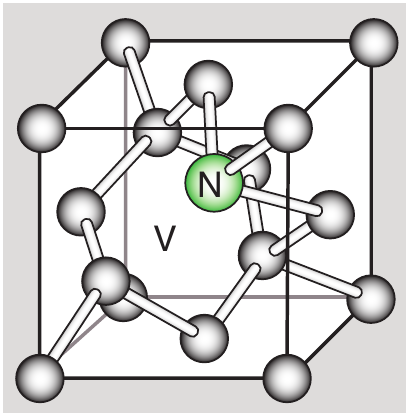}
\caption{The physical structure of the nitrogen-vacancy defect in diamond.}
\label{nvphysical} 
\end{figure}

The nitrogen-vacancy, or NV defect, pictured in Figure \ref{nvphysical} is the most abundant colour centre (optically active defect) in diamond, and results -- as the name suggests -- from the pairing of a vacancy with a substitutional nitrogen on adjacent lattice sites along one of the four [111] crystal axes. The centre acts as an electron trap, and it is the negatively charged species, \nvminus, that is of interest here. The six unpaired electrons are highly localized and well isolated from interactions with the lattice.

\subsubsection{Triplet spin qubit}
\begin{figure}[h]
\includegraphics[width = 1 \columnwidth]{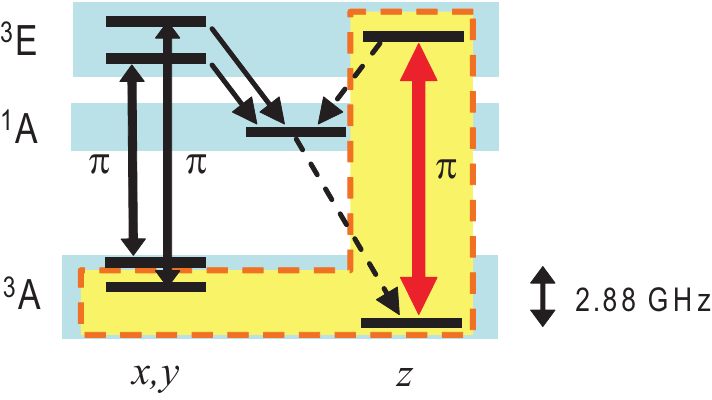}
\caption{ Schematic of the electronic structure of the negatively charged nitrogen-vacancy (\nvminus) defect. Each of the electronic sublevels is separated horizontally into the different spin components. Two of the three spin directions in the $^3$A level are used as the spin qubit. Vertical red arrows show the electric dipole transitions, all of which have linear ($\pi$) polarization. Note that the $^3$E excited state splits into two spin triplets under an electric field transverse to the NV axis \cite{Manson06,Tamarat08}.}
\label{nvelectronicstructure}
\end{figure}

The electronic structure of the \nvminus defect has been established by a number of important works over the past thirty years \cite{davies76,VanOort88,Redman91,Manson06,Tamarat08,Larsson08,Rogers08}, and is shown schematically in Figure \ref{nvelectronicstructure}. The ground state configuration is a \tripletA spin triplet with the $m_s=0$ ($z$) and $m_s=\pm1$  ($x,y$) states split by 2.88 GHz due to a spin-spin interaction \cite{VanOort88,Redman91}. The electron spin in the \tripletA manifold couples to a triplet excited state \tripletE via linearly polarized electric dipole transitions of 1.945 eV (637 nm) that preserve spin under the C$_{3v}$ symmetry of the defect \cite{davies76}. Two singlets \singletA and $^1$E reside at energies between the \tripletA and \tripletE manifolds and are responsible for intersystem crossing transitions between the triplet spin projections \cite{Manson06,Rogers08}. The shaded `L' region highlights the levels that might be used for a measurement-based entanglement scheme of the type outlined in figure \ref{erasure_fig}.

\begin{figure}[h]
\centering
\includegraphics[width = 0.95 \columnwidth]{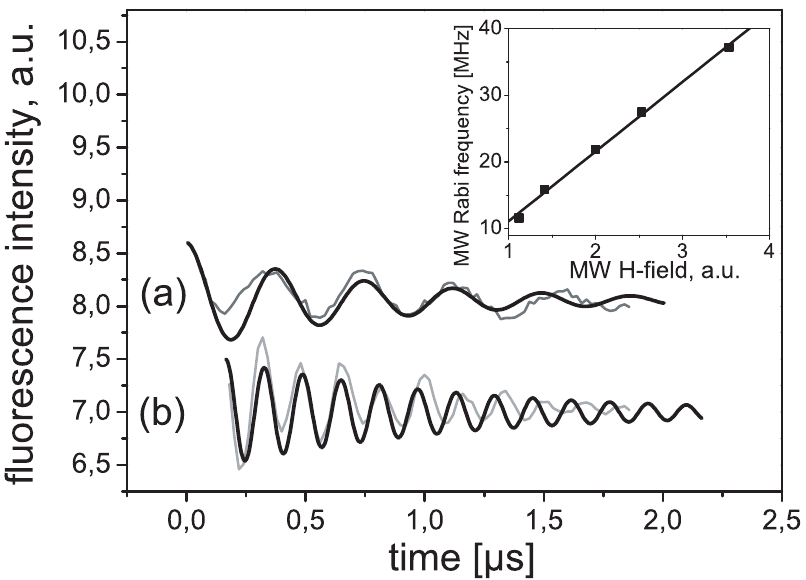}
\caption{The first optically detected Rabi oscillations of single electron spins in a single \nvminus colour centre. Data are shown for two values of the microwave Rabi frequency; curve (a) corresponds to a slow Rabi frequency (approximately 16 MHz), and curve (b) corresponds to a faster Rabi frequency (39 MHz). The solid grey lines represent the measured data, and the thick lines show the simulations based on the microwave optical Bloch equations. The inset shows the linear dependence of the observed modulation frequency on the applied microwave field amplitude, thus confirming the origin of the observed oscillations. The decoherence times in these data are of order a microsecond -- decoherence times of order a millisecond have since been recorded in higher purity material. Reprinted with permission from \cite{Jelezko04}. Copyright (2004) by the American Physical Society.}
\label{nvspinrabi} 
\end{figure}

Following the first optically detected magnetic resonance experiments on single NV centres reported by Gruber et al in 1998 \cite{Gruber97}, it was the ability to polarize and measure single spin nutations, reported by Jelezko {\em et al} in 2004 \cite{Jelezko04}, that ignited the recent interest in using \nvminus centres for quantum information applications. The mechanisms by which polarization and measurement operate hinge on the role of the intermediate singlet manifold, which is populated by relaxation preferentially from the \tripletE$_{x,y}$ levels and depopulates preferentially into the \tripletA$_z$ level. As a result of these asymmetries, the intensity of the \tripletE to \tripletA luminescence under a brief non-resonant excitation provides a measure of the spin $z$ population, whilst after prolonged non-resonant excitation the centre is highly polarized in favour of spin $z$. By polarising the spin, applying a microwave pulse tuned to the \tripletA splitting, and then measuring the spin, Jelezko {\em et al} measured the Rabi oscillations as a function of the microwave pulse duration shown in Figure~\ref{nvspinrabi}.  
 
The Rabi oscillations in Fig. \ref{nvspinrabi} decay with a time constant of order a microsecond. This time scale corresponds to the dephasing time $T_2$ of the electron spin resulting from fluctuating electron spins on nearby substitutional nitrogen atoms. In higher purity materials, dephasing times approaching a millisecond have since been observed at room temperature \cite{Gaebel06} which are presumed to be limited by the nuclear spin bath provided by the \cthirteen isotopes. Much longer dephasing times are anticipated in isotopically purified \ctwelve ~ material that are currently being developed \cite{EQUIND}. As well as spin manipulation with an appied microwave field, Santori {\em et al} have demonstrated optical manipulation of the spin state using coherent population trapping via a \tripletE state of mixed spin character \cite{Santori06}, a capability that may bring benefits in spatially localized spin control. It is not yet clear however whether arbitrary rotations of the spin qubit can be performed in this manner. 

\subsubsection{Brokering}
Jelezko's 2004 paper also reported hyperfine splitting of the \tripletA sublevels under the application of a small static magnetic field, and several works have since demonstrated coupling of the \nvminus electron spin with other neighbouring spins. This coupling is both a source of dephasing and a potentially powerful resource. David Awschalom's group at the University of California at Santa Barbara have studied coupling to the electron spins on nearby substitutional nitrogen atoms \cite{Epstein05,HansonPRL06,HansonScience08}, whilst the Stuttgart and Harvard groups have demonstrated coupling to the nuclear spin of \cthirteen \cite{Jelezko04a,Childress06}. In their 2007 Science paper \cite{Dutt07}, Dutt {\em et al} made use of the Larmor precession of the \cthirteen nuclear spin in a small applied magnetic field to demonstrate high fidelity transfer of quantum information from the \nvminus electron spin to the \cthirteen nuclear spin and back again. They measured no decay of a Hahn echo signal on a time scale of 20 ms, and inferred from this a dephasing time for the \cthirteen nucleus of at least a second. With a hyperfine splitting of order 100 MHz resulting from the \cthirteen nucleus occupying one of the three sites adjacent to the vacancy in the \nvminus, a two qubit gate is in principle achievable in just a few nanoseconds. Such a process is ideal for the kind of brokering scheme described in section \ref{growing}, and illustrates the huge potential of diamond as a host material for spin qubits. As also noted in section \ref{growing}, further advantage can be gained by increasing the number of qubits per node, and this can be achieved by identifying NV's that are coupled to multiple \cthirteen nuclear spins, such as the one used by Neumann {\em et al} to demonstrate a Greenberg Horne Zeilinger (GHZ) cluster state produced using the \nvminus electron coupled to three nuclei \cite{Neumann08}.

\subsubsection{Path erasure}

For measurement-based entanglement to be successful, a suitable optical excitation/emission route must be established, and with this in mind recent attention has turned to the properties of the \tripletE manifold. The primary tool for probing these levels is photoluminescence excitation (PLE) spectroscopy, which takes advantage of the fact that at cryogenic temperatures, 96$\%$ of photons emitted reside in broad phonon sidebands to the resonant zero phonon line (ZPL). By monitoring the intensity of the phonon sideband emission as a narrow line width laser is tuned through resonance with the ZPL, the energies and widths of the excited states can be characterized \cite{Tamarat06,Tamarat08}. In doing so is generally necessary to artificially shorten the $T_1$ time of the spin to prevent spin pumping from quenching the PLE signal, and to `repump' the centre with green of blue light to correct for ionization.

\begin{figure}[h]
\centering 
\includegraphics[width = 0.95 \columnwidth]{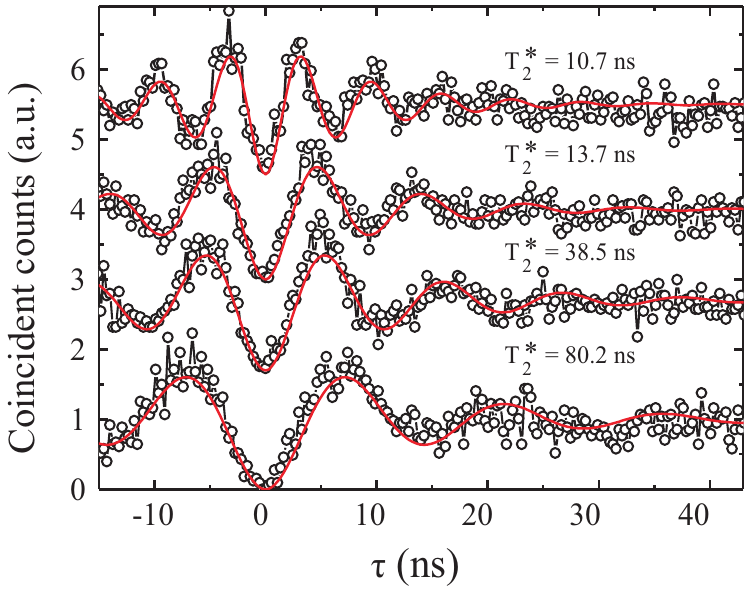}
\caption{ Second order fluorescence intensity correlation function for the NV centre at low temperature under resonant excitation of the spin $z$ optical transition. The different data sets correspond to different excitation laser intensities, giving Rabi frequencies of 0.44, 0.59, 0.67 and 1.00 GHz (increase from bottom to top) The solid red line is an analytic fit to a solution of the optical Bloch equations for a damped two level system. Reprinted with permission from \cite{Batalov08}. Copyright (2008) by the American Physical Society.}
\label{nvopticalrabi}
\end{figure}

Spin measurement for measurement-based entanglement will require spectral selection of the required transition, pulsed excitation, and detection of the spontaneously emitted photon. To be confident that the electronic state reverts to the original spin after measurement it is necessary to select a highly cyclic transition, which under most conditions is the spin $z$ as indicated in figure \ref{nvelectronicstructure}. To be confident that no unknown {\em phase} has been imparted on the quantum state by the measurement, the energetic width of the excited state must be equal to the spontaneous emission rate, that is to say that $T_2^{pdp} >> T_1$ in the \tripletE level. Such conditions have indeed been observed in selected diamond materials \cite{Tamarat06,Batalov08}, and are very close to being manufactured in new ultra-pure synthetic diamond. Figure \ref{nvopticalrabi}, from ref.\cite{Batalov08}, shows the second order temporal correlation of photons emitted under continuous resonant excitation, and reveals clear Rabi oscillation of the excited state population in the wings of the antibuching dip at $\tau=0$. Pure dephasing times as long as 80 ns were recorded and, encouragingly, the dephasing rate measured is approximately proportional to the excitation intensity (ie to the square of the optical Rabi frequency), suggesting that significant further increases may be obtained under short pulsed excitation. To achieve path erasure between two NV's it will additionally be necessary to bring the ZPLs into resonance with each other, probably using a Stark shift \cite{Tamarat06}, and to excite and detect resonantly with the optical transition. The latter remains an outstanding challenge that will be discussed in more detail in the next section.  

\subsubsection{Measurement efficiency}

\begin{figure}[h]
\centering
\includegraphics[width = 0.95 \columnwidth]{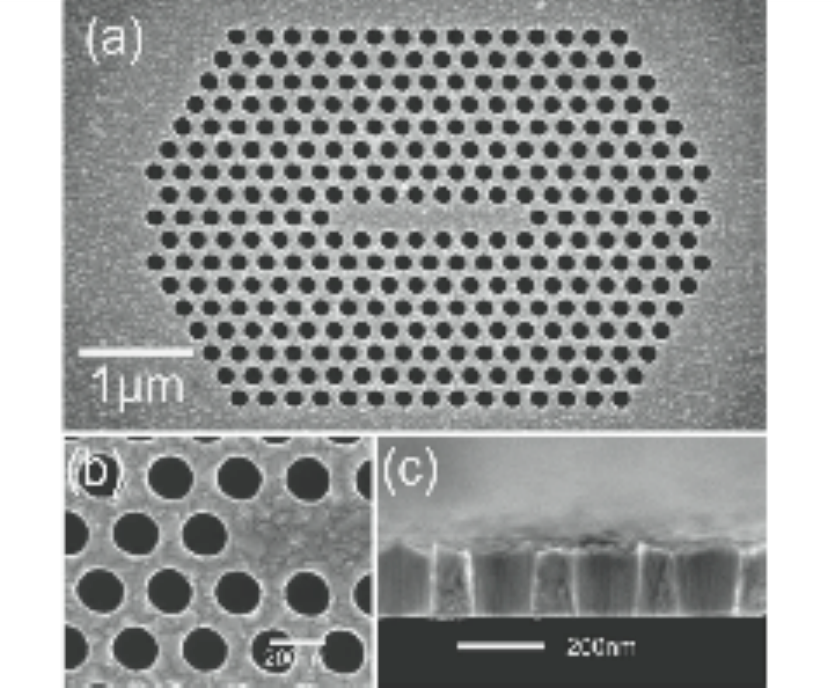}
\caption{(a) A scanning electron micrograph of a suspended L7 photonic crystal cavity in polycrystalline diamond with hole separation 240 nm and hole radius 80 nm. (b) An enlarged picture of air holes. (c) A cross sectional image of the air holes. The thickness of the membrane is about 160 nm. The sidewall of the holes is tilted by 3$\deg$. Reused with permission from \cite{Wang07}. Copyright 2007, American Institute of Physics.}
\label{nvpccavity} 
\end{figure}

That only 4$\%$ of optical emission from the \tripletE states is into the ZPL is inconvenient, since photons in the phonon sideband can not be used in the kinds of entanglement operations described in section \ref{entanglement}; phonon emission would provide a way for `nature' to know which of the NV's had emitted a photon and would therefore eliminate path erasure. One solution is the use of an optical cavity to enhance emission into the ZPL through the Purcell effect \cite{Purcell46}. Su {\em et al} \cite{Su08} have solved the master equation for emission from a single \nvminus coupled resonantly into a leaky cavity with mode volume $V = \lambda^3$, where $\lambda$ is the wavelength of the ZPL emission in the diamond, and found that a quality factor of $Q = 10^4$ would suffice to couple more than 95$\%$ of emission into the ZPL. According to Tomljenovic-Hanic {\em et al} \cite{Tomljenovic06} such parameters are realistic targets for photonic crystal cavities in diamond like the one shown in Figure \ref{nvpccavity}, produced by Evelyn Hu's group in Santa Barbara. A diamond photonic chip can therefore be envisaged in which optical cavities allow efficient coupling of the zero phonon transition with planar photonic crystal waveguides, and in which the path erasure is built in through use of photonic crystal beam splitters. As an alternative to photonic crystal cavities, which may take some time to optimize, the fact that the Purcell enhancement factor is proportional to $Q/V$ means that a larger Fabry Perot cavity with a higher Q factor could achieve the same effect and enable efficient coupling to free space optics.

\subsubsection{Implementation}
A first implementation of an NV centre based quantum computer would probably involve a single centre in each cavity, so as to maximise the ability to perform simultaneous control over spins, photons, and the local environment of the centre. Deterministic entanglement between the electron spin of the defect and nearby nuclear spins would allow a brokering scheme to be operated, and would provide the long coherence times required for scalability. A reasonable estimate for the time required to perform an entanglement operation is 200 ns. Combined with a modest measurement efficiency of only $\eta = 0.01$, giving an EO efficiency of $0.5\eta^2 \simeq 5\times10^{-5}$, this would allow the creation of a graph state edge every 4 ms, within which time the decoherence of other nuclear spins would be well within the suggested fault tolerances for cluster state QIP \cite{raussendorf07a}. The principal source of errors is likely to be from residual instabilities in the optical transition, and maintaining path erasure between pairs of qubits will be challenging; this difficulty is not fundamental however, and should be lessened or eliminated by further development of the diamond material. The coherent properties of spins in diamond also open up the enticing possibility of room temperature operation -- the challenge here would be to develop strongly coupled microcavities that would enable a STIRAP type readout scheme to be achieved, thereby avoiding population of the broadened \tripletE state.

\subsection{Semiconductor quantum dots}
\label{QD}
\begin{figure}[h]
\centering
\includegraphics[width = 0.9 \columnwidth]{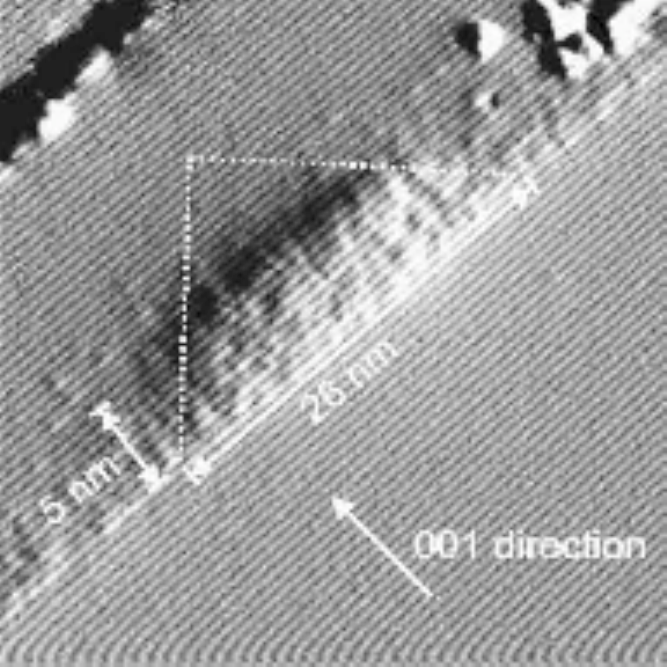}
\caption{40$\times$40 nm cross-sectional scanning tunneling micrograph of a cleaved InAs quantum dot grown on [001] GaAs by the Stranski Krastanow method. The indium rich region has a larger lattice constant than does the surrounding GaAs, and so bulges upwards providing a clear picture of the extend of the quantum dot. In the upper right corner some cleavage debris is visible. Reused with permission from \cite{Bruls02}. Copyright 2002, American Institute of Physics.}
\label{qdxstm} 
\end{figure}

Quantum dots provide an alternative means to isolate single electrons from the decohering effects of bulk materials and present the substantial advantages over crystal defects that their properties can be engineered to a greater degree, and they can more easily be integrated into devices. Dots that confine both electrons and holes in the same region of space possess strong atom-like optical transitions between the valence and conduction bands that can be used for optical measurement of spin qubits. The most well studied species is the self-assembled or Stranski Krastanow quantum dot (SKQD), which results from a lattice mismatch between epitaxial layers of III-V and II-VI semiconductors. The spontaneous island formation that occurs during growth provides a highly localized reduction in the band gap surrounded by a continuous defect-free lattice (Figure \ref{qdxstm}). 

\subsubsection{Electron and hole spin qubits in quantum dots}
The inter-particle Coulomb energy $U$ in self-assembled quantum dots is of order tens of meV, so that at temperatures of up to about 10K $k_BT<<U$ and single electrons or holes can be added and removed with fine control using a simple field effect structure \cite{Warburton00}. It is therefore possible to isolate single carriers in the dots, and to use the spin of the particle as a qubit. As we have noted in section \ref{decoherencespins}, the spin coherence time of electrons in self-assembled quantum dots in III-V semiconductors is of order 1 $\mu$s limited by a hyperfine interaction with the randomly fluctuating spins of the host nuclei, although polarising the the nuclear spin bath, and use of confined holes rather than electrons \cite{Gerardot08} are possible routes to extending $T_2$ beyond those currenly measured. $T_1$ times can be of order tens millieconds \cite{kroutvar04} but may be shorter if either the electron spin states are degenerate to within the hyperfine line width \cite{Muller07}, or if interaction with a nearby Fermi sea of electrons allows a fast `co-tunneling' process \cite{Smith05}.

\begin{figure}[h]
\centering
\includegraphics[width = 1 \columnwidth]{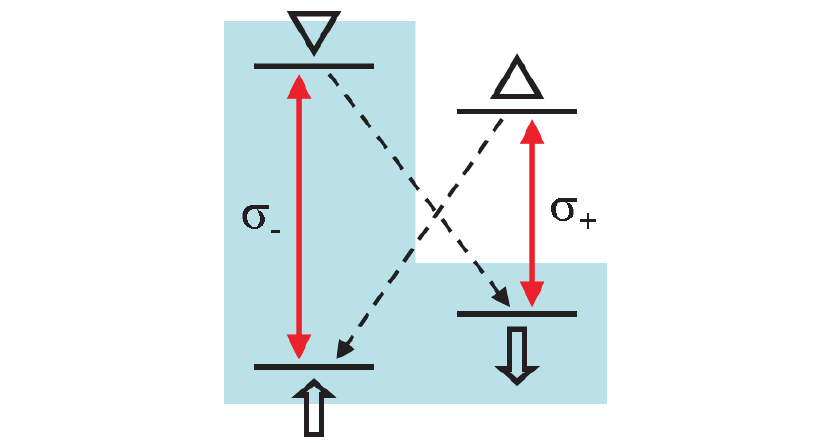}
\caption{The electronic structure of a self-assembled quantum dot charged with a single electron, and with a small applied magnetic field that Zeeman splits the otherwise degenerate spin states. Optical transitions from the electron states with different spins (up/down arrows) couple to the trion excited states (up/down triangles) with circularly polarized transitions (red double arrows), while weak spin flip Raman processes (diagonal dashed arrows) provide a means for pumping the spin state.}
\label{qdelectronicstructure} 
\end{figure}
 
A simplified energy level diagram for a quantum dot with a single trapped electron is shown in figure \ref{qdelectronicstructure} (it is easy to see how this diagram should be adapted for holes). The spin degeneracy of the $1s$ ground state has been lifted by applying a weak B-field in the Faraday orientation (parallel to the optical axis). Optical transitions of opposite circular polarizations couple the two lower energy states to their respective trion states in which the two electron spins (arrows) form a singlet and the angular momentum is determined only by the hole (triangle). 

As with NV centres, prolonged resonant excitation of one of these optical transition results in spin pumping to the off-resonance spin state, this time via a weak spin-flip Raman processes (diagonal dotted lines in the figure) -- a process that is potentially useful as it provides an efficient, although relatively slow, means to initialize the spin state \cite{Atature06}. Faster spin initialisation can be achieved using a magnetic field in the Voigt geometry which mixes the excited state spin character \cite{XuPRL07}. The first optically detected spin resonance experiments have recently been reported, which rely on the fact that a microwave field resonant with the Zeeman splitting of the electron spin states can be used to reduce $T_1$ and thereby restore an absorption signal that would otherwise be quenched by spin pumping \cite{Kroner08}. Controlled single spin rotations have yet to be demonstrated however, although the rich electronic structure of the valence band in these materials offers opportunities for spin manipulation using fast optical pulses \cite{Berezovsky08}, which might be preferred over the use of slower and spatially delocalised rf fields. 

\subsubsection{Spin measurement}
 The great strength of self-assembled quantum dots for QIP purposes lies in their optical transitions. The transition dipoles are typically about twenty times larger than in single atoms or point defects since the electron and hole wave functions extend over $\sim10^4$ lattice sites, therefore allowing coherent control of the transition to be achieved with electromagnetic field intensities that are of order four hundred times lower. Moreover the optical transitions have polarizations that depend on the carrier spin, making for a convenient means of performing spin selective operations without the need for fine spectral discrimination, and they do not couple strongly to optical phonons, so that the vast majority of the total oscillator strength is in the zero phonon line. These strong, narrow transitions have enabled numerous quantum optics experiments to be performed such as optical Rabi flopping \cite{Zrenner02}, state dressing \cite{XuScience07}, and resonance fluorescence \cite{Vamivakas08}. In figure \ref{qdelectronicstructure} an $L$ configuration of levels suitable for MBQC is identified that utilises the two electron spin orientations and the trion with the hole in the spin up state ($J=3/2$), whereby the optical transition is with a right hand polarized photon. Resonance fluorescence of the optical transition would then provide the required spin measurement. Resonance fluorescence of a neutral dot in a cavity was recently reported by Muller et al \cite{Muller07} using an attractive approach that permits efficient collection of the emitted light into a resonant cavity mode that is spatially decoupled from the excitation. Adapting such a scheme to a charged dot would approximate the necessary spin measurement conditions for entanglement operations to be realised.

\subsubsection{Path erasure}
In 2003, Santori et al showed that successive excitations of a single quantum dot at at temperature of 4K could produce indistinguishable photons \cite{Santori02}, thus demonstrating that dephasing is slow compared with the exciton lifetime and that spectral drift is slow compared with the 2 ns time delay between the excitation pulses. Indistinguishability between photons from different dots has yet to be achieved however, due to fluctuations in the transition energies on millisecond time scales. Since no phonon sidebands are visible in the emission from SKQDs, PLE experiments can not be used to characterise the dephasing rate in the lowest energy exciton states, and absorption spectroscopy \cite{Seidl05,Hogele05} or resonance fluorescence \cite{Muller07,Vamivakas08} are required. The narrowest line widths measured by these methods are about twice that of the lifetime limited value \cite{Seidl05,Hogele05}. Absorption features in p-type doped field effect structures measured to date are substantially broader than in n-type doped structures \cite{Gerardot08} suggesting that the source of the spectral drift could be the presence of local lattice impurities or defects. It is therefore likely that strictly lifetime limited line widths can be achieved with some development of the materials growth process.

\begin{figure}[h]
\centering
\includegraphics[width = 1 \columnwidth]{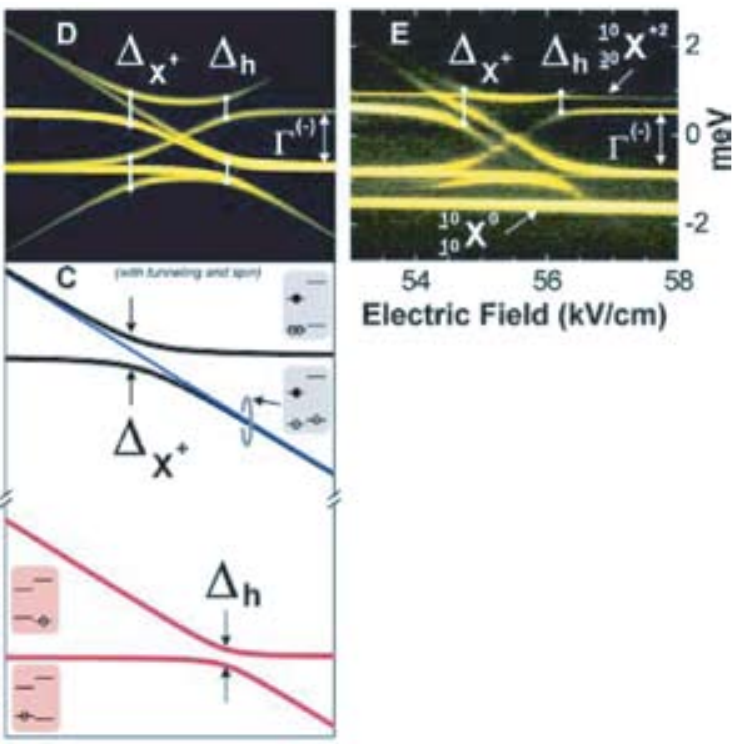}
\caption{Anticrossing between electron and hole states in a quantum dot molecule. Plot C shows the energy level scheme as a function of applied bias for the hole-only states (red lines) and the trion states (black lines). Plot D shows the predicted energies and strengths of the optical transitions resulting from these energy levels, and Plot E shows the corresponding experimental results. From \cite{Stinaff06}. Reprinted with permission from AAAS.}
\label{qdanticross} 
\end{figure} 

\subsubsection{Brokering}
The best prospect for implementing a brokering scheme using SKQDs lies in the use of pairs of dots that can be selectively coupled by some control field or biasing of the device. This is familiar territory for experiments using quantum dots formed by electrical gating of high mobility two dimensional electron gases in GaAs \cite{petta05}, but such dots do not possess suitable optical transitions and require millikelvin temperatures due small quantization energies of typically about 1 meV. Growth of SKQDs in close proximity can be achieved by the vertical stacking effect that is observed when a second layer of dots is grown on top of the first. Coupling between dots has been demonstrated via tunneling of electrons \cite{Robledo08} and holes \cite{Stinaff06}, whereby an applied electrical bias tunes the two dots through the tunneling resonance and anti-crossing of the quantum states is observed. Figure \ref{qdanticross} shows a particularly nice example from ref. \cite{Stinaff06} in which anti-crossings in both the hole-only and the trion state lead to photoluminescence lines that reveal the two respective tunneling energies $\Delta_h$ and $\Delta_{X^+}$. Combined with the spin-polarization selection rules for trion formation, it is not difficult to see how such a system can be used to perform controlled quantum gates conditional on the spin state of the hole. Schemes have also been suggested involving F\"{o}rster energy transfer \cite{Nazir04} instead of tunneling, but these may be slightly more challenging to demonstrate since the relative exciton energies of stacked quantum dots is only adjustable over a very small range. 

\subsubsection{Measurement efficiency}
\begin{figure}[!h]
\centering
\includegraphics[width = 1 \columnwidth]{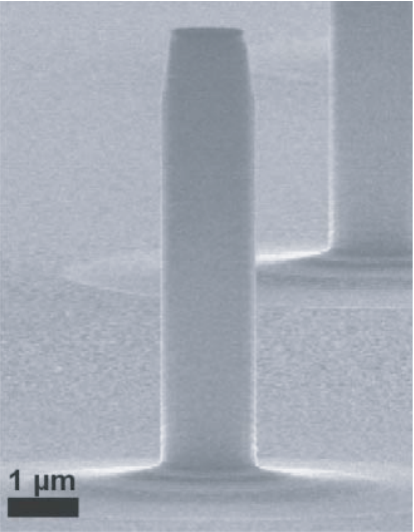}
\caption{SEM image of a monolithic micropillar structure containing a single layer of InAs Stranski Krastanow quantum dots between two highly reflective GaAs/AlAs Bragg reflectors. Reproduced with permission from \cite{Press07}, copyright the American Physical Society 2007}
\label{qdmicropillar} 
\end{figure}

Self-assembled quantum dots have been integrated into cavities of a variety of geometries, demonstrating both large Purcell enhancements of the spontaneous emission rate \cite{Gerard98,Press07} and strong coupling to form cavity polaritons \cite{Reithmaier04,Badolato05,Press07}. The best geometry for high collection efficiency is the Fabry Perot, and micropillar structures such as the one shown in figure \ref{qdmicropillar} are now readily produced. Photon collection efficiencies as high as $\simeq 0.97$ have been claimed \cite{Press07}. 

\subsubsection{Implementation} 
A pair of coupled quantum dots, each doped with a single electron or hole, situated in a pillar microcavity and manipulated by applied dc and poissibly rf fields, could thus provide a suitable module for the quantum computing scheme envisaged. Operation would almost certainly be at liquid helium temperatures to minimise phonon induced dephasing; the main source of error would then be decoherence due to effects of the nuclear spin bath, and the potential of these qubits very much depends on to what extent such decoherence can be suppressed. If spin rotations can be achieved using fast laser pulses, the time for an entanglement operation would be determined primarily by the spontaneous emission steps and may ultimately be as short as a nanosecond. High efficiency cavity coupling and photon detection may provide $\eta$ as large as 0.5 whereby a single graph edge could be generated in just 10 ns.

\section{Outlook}
\label{outlook}

With an effective brokering scheme in place the prospects of constructing large cluster states using \nvminus appear good. The main challenges for \nvminus then are to produce synthetic material which contains centres that emit indistinguishable photons, and to construct microcavity devices with which to exercise control over the spontaneous emission process. Some effort is focused on identifying new spin active colour centres in diamond with luminescence that is less strongly coupled to phonons. The so-called NE8 centre, a nickel atom surrounded by four interstitial nitrogen atoms, is one candidate \cite{Gaebel04} that has been identified optically, but a clear spin signature has yet to be observed. In the absence of a new defect being identified, the \nvminus appears to possess all the attributes necessary for pursuing scalable MBQC. 

The stronger, polarization sensitive optical transitions of quantum dots makes spin measurement compatible with MBQC both easier and considerably faster than in \nvminus, but these gains are at the cost of faster spin decoherence. Even with high measurement efficiencies, creating (and demonstrating) multipartite entanglement on a microsecond time scale appears rather daunting. Development of tricks that negate the influence of the nuclear spin bath will go a long way to improving the prospects of achieving scalable MBQC using quantum dots, and this is a clear focus of much of the current research.

Whilst it is likely to be easier at first to construct a physically scalable computer that comprises a distributed network of remote nodes as indicated in the title figure, scalability in a manufacturing sense requires mass production. An ultimate goal is therefore to fabricate fully integrated chips containing cavities, waveguides, optical switches and beam splitters, with sufficient redundancy to accommodate malfunctioning qubits. Certainly to fabricate such structures from either semiconductors or from diamond will require substantial process development. Nevertheless, the challenges for scalable quantum computing in the solid state appear more tractable in the short term using measurement-based protocols rather than traditional circuit-based schemes.

%%% Give an acknowledgement if you have to thank someone, are 
%%% supported by someone or something like that:
\begin{acknowledgement}
SCB and BWL would like to thank the Royal Society for financial support. This work is also supported by the National Research  
Foundation and Ministry of Education, Singapore.
\end{acknowledgement}

%%% Give your bibliography:


\begin{thebibliography}{100}
\bibitem{EQUIND} These materials are being developed as part of the European Commission Framework 6 EQUIND project.
\end{thebibliography}
\end{document}